\documentclass{emulateapj}
\usepackage[latin1]{inputenc}
\usepackage{amsmath}
\usepackage{amsfonts}
\usepackage{amssymb}
\usepackage{graphicx}
\usepackage{natbib}
\usepackage{makecell}
\usepackage{epstopdf}
\usepackage{tabularx}
\usepackage{booktabs}
\usepackage{threeparttable}

\begin{document}
	\title{Thermally Unstable Cooling Stimulated by Uplift: The Spoiler Clusters}

	\author{C.~G. Martz$^{1,2,\ast}$}
	\author{B.~R. McNamara$^{1,2,3}$}
	\author{P.~E.~J. Nulsen$^{5,6}$}
	\author{A.~N. Vantyghem$^{7}$}
	\author{M-J. Gingras$^{1,2}$}
	\author{Iu.~V. Babyk$^{1,2,4}$}
	\author{H.~R. Russell$^{8}$}
	\author{A.~C. Edge $^{9}$}
	\author{M. McDonald$^{10}$}
	\author{P.~D. Tamhane$^{1,2}$}
	\author{A.~C. Fabian$^{11}$}
	\author{M.~T. Hogan$^{1,2,3}$}

	\affil{
		$^{1}$ Department of Physics and Astronomy, University of Waterloo, 200 University Avenue West, Waterloo, ON, N2L 3G1, Canada\\
		$^{2}$ Waterloo Centre for Astrophysics, University of Waterloo, 200 University Avenue West, Waterloo, ON, N2L 3G1, Canada \\
		$^{3}$ Perimeter Institute for Theoretical Physics, 31 Caroline St N, Waterloo, ON, N2L 2Y5, Canada \\
		$^{4}$ Main Astronomical Observatory of NAS of Ukraine, 27 Academica Zabolotnogo St, 03143, Kyiv, Ukraine\\
		$^{5}$ Harvard-Smithsonian Center for Astrophysics, 60 Garden Str, Cambridge, MA 02138, USA \\
		$^{6}$ ICRAR, University of Western Australia, 35 Stirling Hwy, Crawley, WA 6009, Australia \\
		$^{7}$ University of Manitoba,  Department of Physics and Astronomy, Winnipeg, MB R3T 2N2, Canada \\
		$^{8}$ School of Physics \& Astronomy, University of Nottingham, University Park, Nottingham, NG7 2RD, United Kingdom \\
		$^{9}$ Department of Physics, University of Durham, South Road, Durham DH1 3LE, United Kingdom \\
		$^{10}$ Kavli Institute for Astrophysics and Space Research, Massachusetts Institute of Technology, 77 Massachusetts Avenue, Cambridge, MA 02139 \\
		$^{11}$ Institute of Astronomy, Madingley Road, Cambridge CB3 0HA, United Kingdom \\
	}
	\altaffiltext{*}{cg2martz@uwaterloo.ca}
	
	\begin{abstract}
     Chandra X-ray observations are analyzed for five galaxy clusters whose atmospheric cooling times, entropy parameters, and cooling time to free-fall time ratios within the central galaxies lie below $1$ Gyr, below $30$ keV cm$^{2}$, and between $20 \lesssim \text{min}(t_{\rm cool}/t_{\rm ff}) \lesssim 50$, respectively. These thermodynamic properties are commonly associated with molecular clouds, bright H$\alpha$ emission, and star formation in central galaxies. However, all have H$\alpha$ luminosities below $10^{40}\rm ~erg~s^{-1}$ in the ACCEPT database. Star formation and molecular gas are absent at the levels seen in other central galaxies with similar atmospheric properties. Only RBS0533 may host a radio/X-ray bubble which are commonly observed in cooling atmospheres. Signatures of uplifted, high metallicity atmospheric gas are absent. Their atmospheres are apparently thermodynamically stable despite the absence of strong nuclear feedback. We suggest that extended filaments of nebular emission and associate molecular clouds are absent at appreciable levels because their central radio sources have failed to lift low entropy atmospheric gas to an altitude where the ratio of the cooling time to the free-fall time falls below unity and the gas becomes thermally unstable.
	\end{abstract}
	
	\section{Introduction}
	\label{Intro}
	The central radiative cooling timescales of galaxy, group, and cluster atmospheres are often shorter than their ages.  As the atmosphere cools and loses pressure support, it is expected to condense into molecular clouds at rates upward of $100 ~\rm M_\odot ~yr^{-1}$ and to form stars. The atmosphere lost to cooling should be replenished by gas moving inward from larger radii in a slow, steady cooling flow \citep{Fabian_1994}.  However, this phenomenon is not observed \citep{Peterson_Fabian_2006}.  
	
	Hot atmospheres are instead globally stable, in hydrostatic and thermal equilibrium. This stability must be maintained by one or more heat sources \citep{Peterson_Fabian_2006, Pizzolato&Soker_2005, Pizzolato_2010}, the most prevalent being mechanical feedback from a radio active galactic nucleus (AGN) \citep{Voit_Donahue_2005, McNamara_2007}. Atmospheric density fluctuation spectra indicate that mild turbulence, presumably driven by rising X-ray bubbles, is able to gently heat atmospheres uniformly over the cooling regions \citep{Zhuravleva_2018}. Thus, cooling flows do not form in stratified atmospheres, despite their relatively short central cooling timescales. Local atmospheric stability may prevail, in part, because their local dynamical timescales are shorter still than their cooling timescales at all altitudes. 
	
	Thermally unstable cooling is expected to occur when the ratio of the cooling timescale to the local free-fall timescale falls near to and below unity \citep{Balbus_Soker_1989, Nulsen_1986, Pizzolato&Soker_2005, McCourt_2011, Prasad_2018, Voit_2015b, Voit_2019}. This condition is never achieved, even at the very centers of clusters where the cooling time falls below one Gyr.  Instead, the ratio $\rm t_{\rm cool}/t_{\rm ff}$ lies above 10 over the entire cooling region, including the center, where the cooling time is shortest \citep{Hogan_2017, Babyk_2018}. This again indicates that atmospheres are largely thermally stable. Nevertheless, the filamentary nebular line emission and star formation observed in dozens of central galaxies \citep{McDonald_2016} indicate that atmospheres may be thermally unstable locally within a largely stable medium \citep{McCourt_2012}.  
	
    Empirically, the hot atmospheres of central clusters and giant elliptical galaxies contain molecular clouds and young stars preferentially when the central atmospheric entropy and cooling timescales lie below K $\lesssim30$ keV cm$^{2}$ and $t_{\rm cool} \lesssim 1.0\times10^{9}$ yr, respectively \citep{Cavagnolo_2008, Rafferty_2008}. Such systems are much more likely to harbor the radio-inflated X-ray cavities that are stabilizing the atmosphere.  Those with central cooling times exceeding $10^{9}$ yr have a much lower incidence of radio emission \citep{Main_2017}. These observations imply that gas supplied by cooling atmospheres is fueling the nuclear activity that is suppressing cooling and sustaining the feedback loop \citep{Churazov_2001, Pizzolato&Soker_2005, Gaspari_2012}.  While the precise conditions under which thermally unstable cooling occurs in these systems are unclear, hot gas in the central regions of cooing atmospheres should eventually cool to low temperatures.

    \cite{Pizzolato&Soker_2005} suggested that condensations of thermally unstable gas may form in the wakes of jets and radio lobes.  Similarly, hydrodynamic simulations of jets advancing into hot atmospheres indicate that uplifted atmospheric gas and the ensuing turbulence may lead to preferential cooling in the wakes of rising X-ray bubbles \citep{Revaz_2008, Li_2014, Brighenti_2015, Voit_2017, Gaspari_2018}.

    ALMA and NOEMA observations of central galaxies have located molecular clouds preferentially in X-ray bubble wakes \citep{Salome_2011,McNamara_2014, Russell_2017, Russell_2019, Olivares_2019}. But it is unclear whether molecular clouds themselves are being lifted outward or whether the molecular clouds are condensing out of low-entropy hot gas lifted behind the bubbles \citep{Salome_2008}.
    
    With the difficulty of lifting high-column density clouds, \citet{McNamara_2016} proposed that thermally unstable cooling is stimulated when a cooling, low-entropy, atmospheric gas parcel is lifted to an altitude where its free-fall time exceeds its cooling time such that $t_{\rm cool}/t_{\rm ff} \lesssim 1$. Feedback then is thought to suppress cooling on large scales while simultaneously stimulating thermally unstable cooling in the vicinity of the bubble, ensuring a self-sustaining feedback loop.  The loop may be stabilized, in part, by star formation which would quickly consume the cooling gas, thus preventing it from over-feeding the black hole.
    
    The ability to lift cooling gas in a galaxy may be key to triggering thermally unstable cooling. If so, cluster centrals with short atmospheric cooling times yet lacking cold clouds may also be devoid of X-ray cavities capable of lifting the low entropy gas \citep{McNamara_2016}. Here we further examine this hypothesis.

	\begin{table*}[ht]
    \centering
    \caption{\textit{Chandra} data used in our analysis}
    \resizebox{1.0\textwidth}{!}{%
    \begin{tabular}{lcccccccc}

    \hline\hline
    Cluster & z & N$_{\rm H}$ & ObsIDs & \multicolumn{2}{c}{Total Exposure (ks)} & \multicolumn{2}{c}{Cluster Center}    \\ 
    & & ($10^{22}$ cm$^{-2}$) & & Raw & Cleaned & RA (J2000) & DEC (J2000)                                        \\    
    & (1) & (2) & (3) & (4) & (5) & (6) & (7)                                                                      \\ \Xhline{2.5\arrayrulewidth}
    \\
    A2029 & 0.0773 & 0.033 & 891, 4977, 6101 & 107.6 & 103.3 & 15:10:56.077 & +05:44:41.05 \\ 
    \\\Xhline{2.5\arrayrulewidth}
    \\
    A2107 & 0.0411 & 0.0445 & 4960 & 35.57 & 34.8 & 15:39:39.043 & +21:46:58.55 \\ 
    \\\Xhline{2.5\arrayrulewidth}
    \\
    A2151 & 0.0366 & 0.0334 & 4996, 19592$^*$, 20086$^*$, 20087$^*$ & 102.8 & 80.2 & 16:04:35.758 & +17:43:18.54 \\
    \\\Xhline{2.5\arrayrulewidth}
    \\
    RBS0533 & 0.0123 & 0.102 & 3186, 3187, 5800, 5801 & 108.6 & 107.9 & 4:19:38.105 & +2:24:35.54 \\
    \\\Xhline{2.5\arrayrulewidth} 
    \\
    RBS0540 & 0.0397 & 0.0786 & 4183, 19593$^*$, 20862$^*$, 20863$^*$ & 64.5 & 61.6 & 4:25:51.300 & -8:33:38.00 \\
    \\
    \hline\hline
    \end{tabular}%
    
    }\begin{flushleft}\textbf{Columns}: (1) redshift, (2) Column density, (3) Observation IDs used for the analysis, (4) Raw combined exposure of the ObsIDs used, (5) useable exposure after data filtering,(6) RA, (7) DEC. Here, the $^*$ denotes the new data obtained for A2151 and RBS0540.\end{flushleft}
    \label{Table:Chandra_Data}
    \end{table*}

    We have identified and analyzed five clusters drawn from the ACCEPT database \citep{Cavagnolo_2008} whose central atmospheric cooling times and central entropy parameters lie below $10^9$ yr and $30 ~\rm keV ~cm^2$, respectively.  Furthermore, their central $t_{\rm cool}/t_{\rm ff}$ lie in the range of $10-30$. 
    Their atmospheric mean temperatures, within 100 kpc of the centre lie between $1.5-8$ keV, and their central densities and pressures span more than a decade, ranging between $0.01-0.1$ cm$^{-3}$ and $10^{-10}-10^{-9}$ erg cm$^{-3}$, respectively. While small, the sample probes a broad range of environment, from groups to rich clusters.
    
    Their atmospheric properties are similar to other cluster and group atmospheres rich in star formation and molecular gas. Yet these clusters are devoid of molecular gas and star formation at levels detected in other systems \citep[]{McDonald_2010,McDonald_2011}.
    The analysis presented here shows that they are also devoid of X-ray bubbles capable of lifting atmospheric gas to an altitude where it is likely to become thermally unstable. This condition may explain, as we consider here,  why these systems exhibit no spatially-extended nebular emission like that seen in Perseus and other clusters. However, it begs the question why they are apparently thermodynamically stable, when a distributed heat source is required to enforce this stability \citep{McCourt_2012}.  Either their atmospheres are being heated without the production of radio bubbles, or they are in a short-lived state.    
    
    Despite the absence of X-ray cavities in four of the five systems, their central galaxies have detectable radio emission, with A2029's being quite powerful.  It is possible that atmospheric ``sloshing'' \citep{Markevitch&Vikhlinin_2007} and/or the radio jets coursing through their atmospheres create mild turbulence capable of heating their atmospheres and temporarily balancing cooling in these systems \citep{Zhuravleva_2018, Voit_2018, Gaspari_2018}. While speculative, it serves to emphasise the interesting predicament these systems represent. We refer to them as ``spoiler" clusters because they fail to exhibit the usual tracers of cold molecular clouds in their central galaxies that most other systems with similar atmospheric properties display.
    
    Throughout this paper we assume a standard \textit{$\Lambda$}CDM cosmology with $H_{0}=70$ km s$^{-1}$ Mpc$^{-1}$, $\Omega_{m}=0.3$, and $\Omega_{\Lambda}=0.7$. All errors are $1\sigma$ unless otherwise stated.

	\begin{figure*}[htb]
		
		\centering
		\includegraphics[height=75mm, width=75mm]{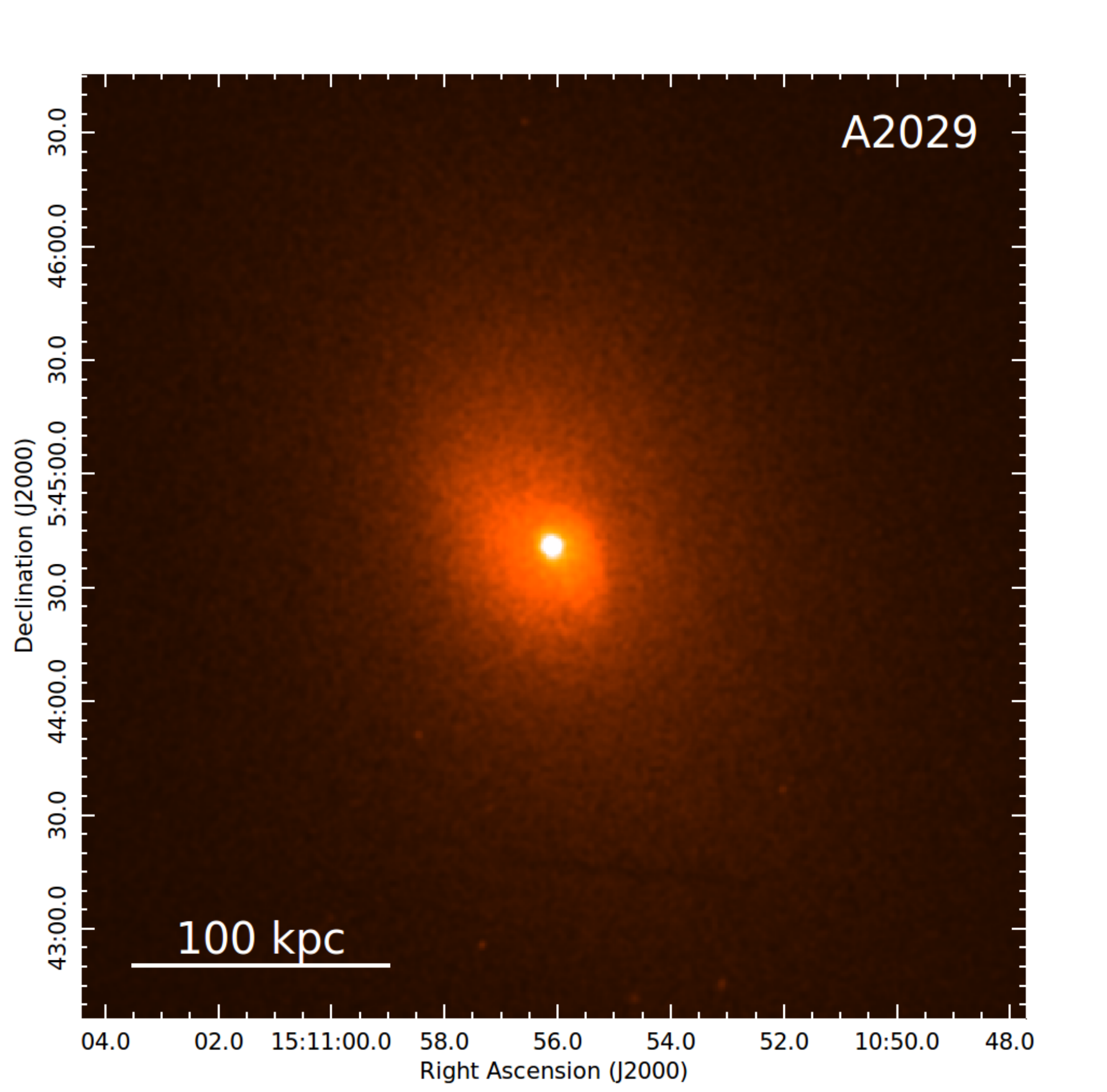}
		\includegraphics[height=75mm, width=75mm]{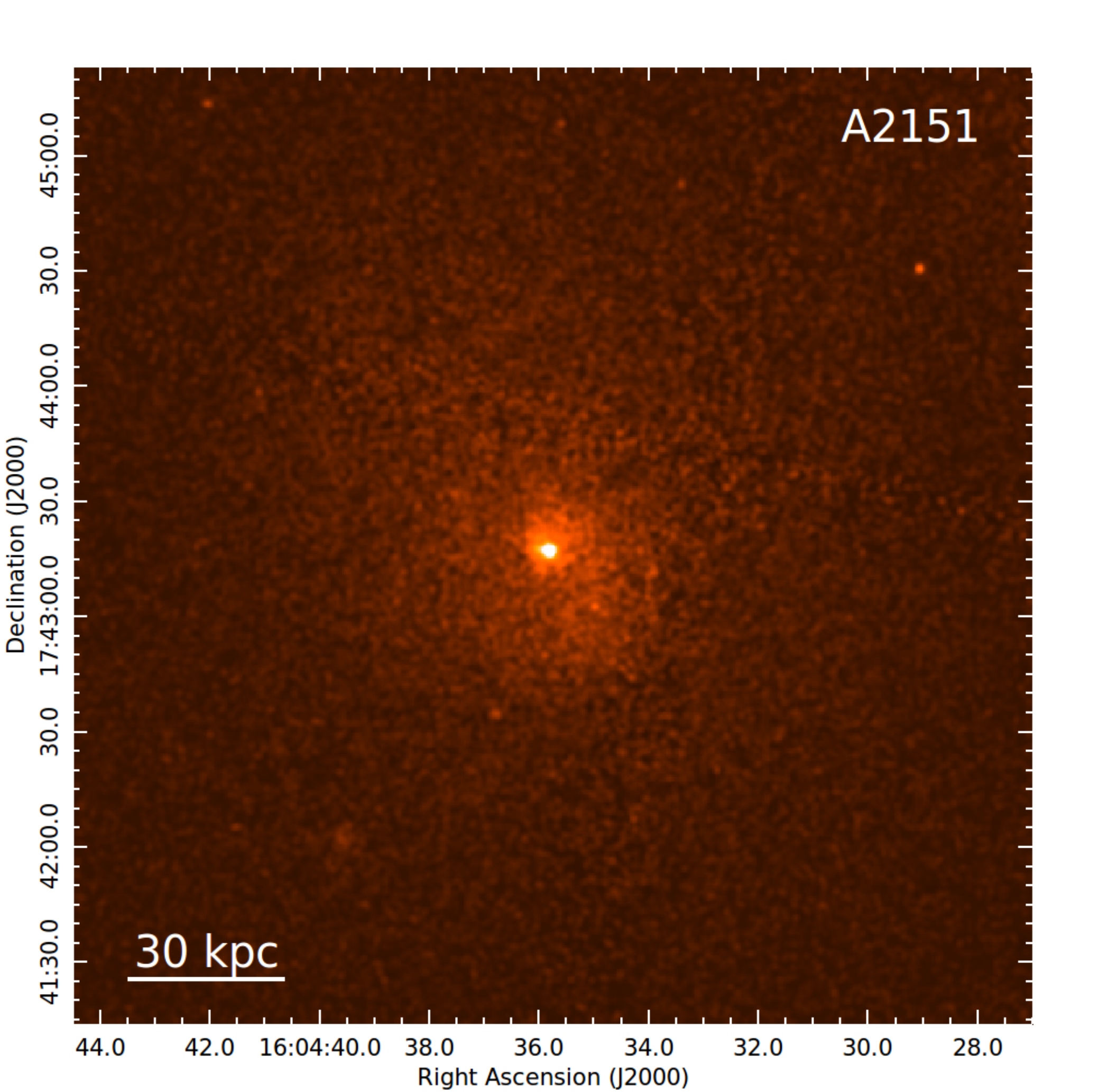} 
		\includegraphics[height=75mm, width=75mm]{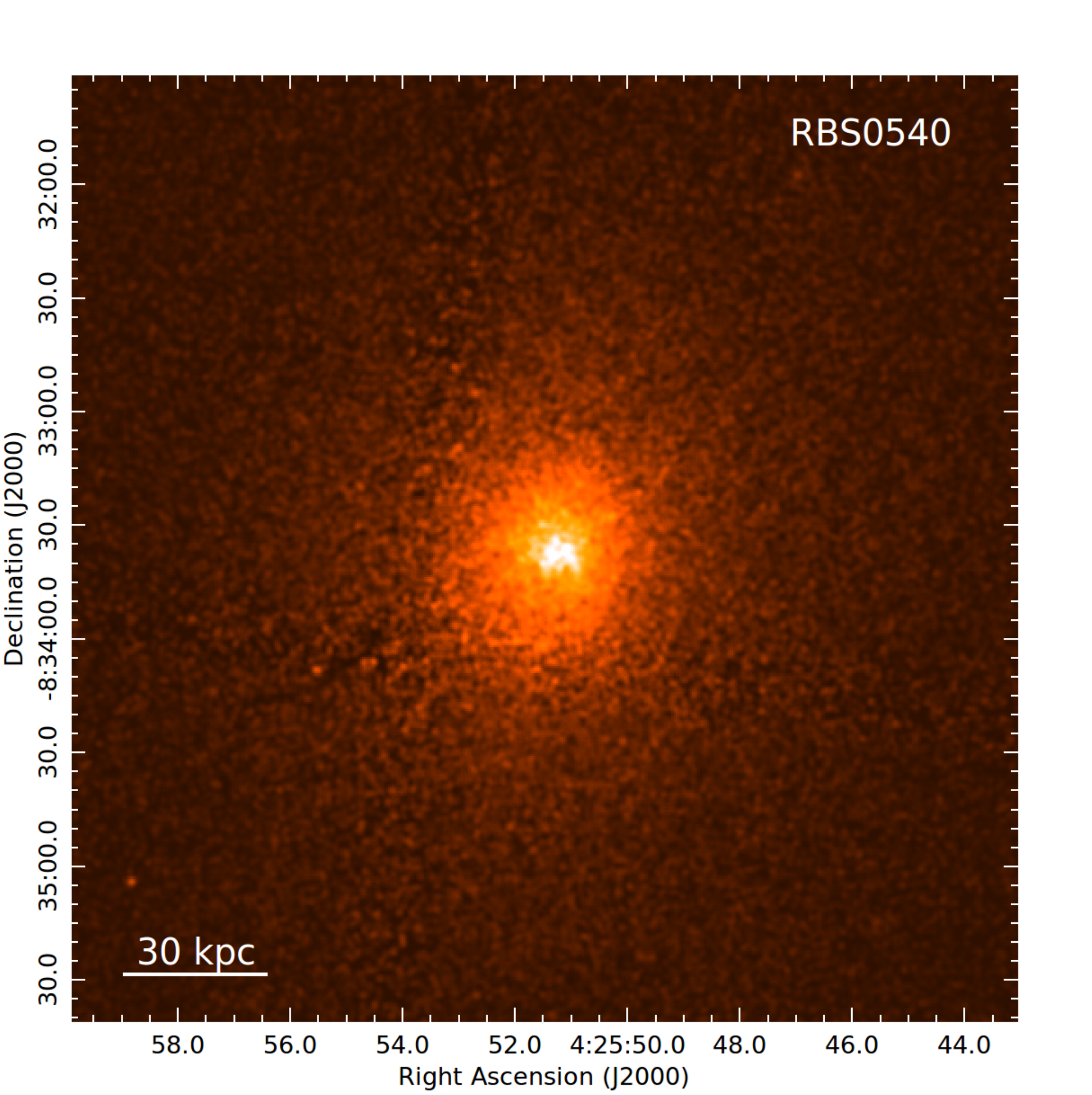}
		\includegraphics[height=75mm, width=75mm]{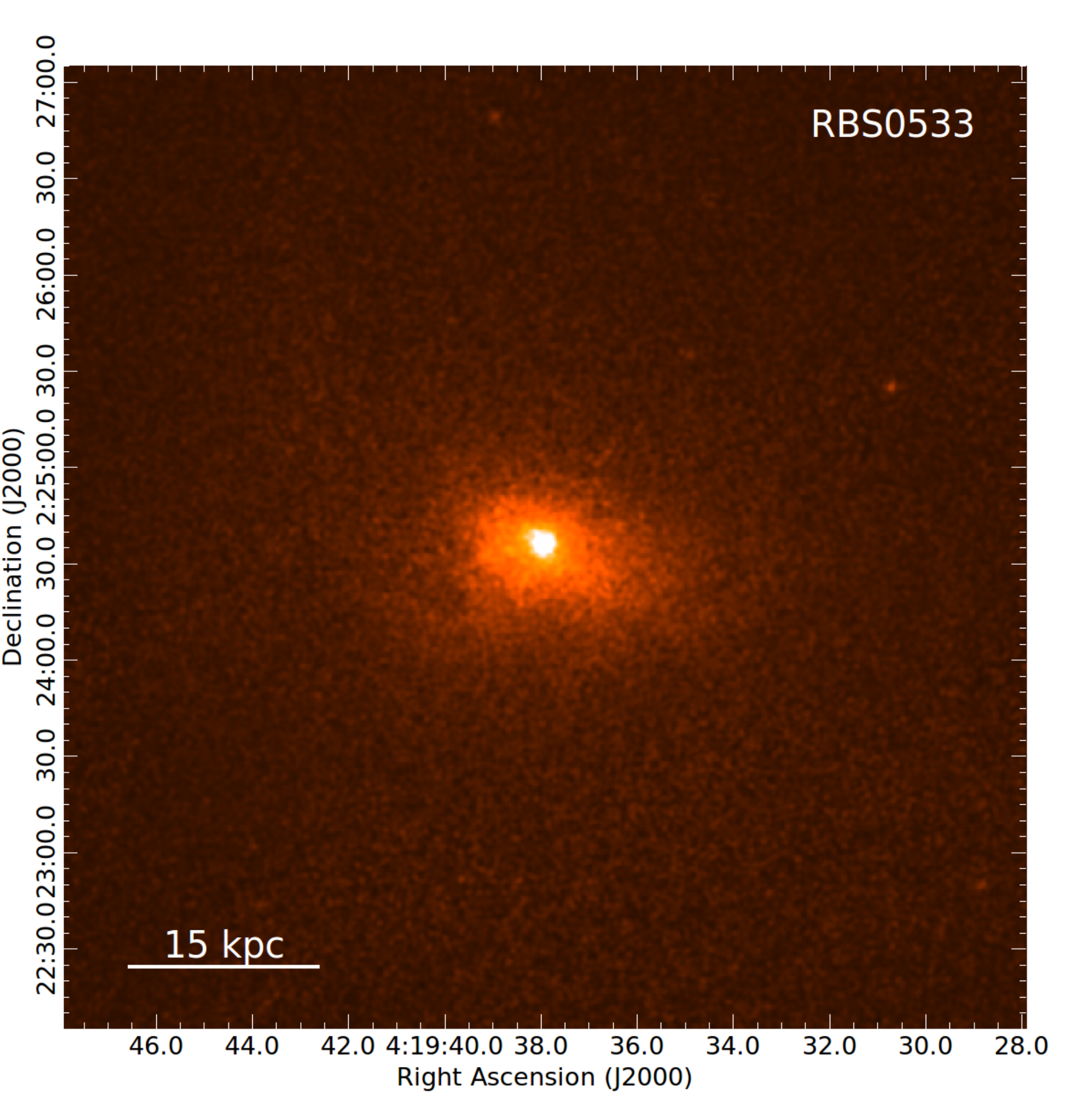}
		\includegraphics[height=75mm, width=75mm]{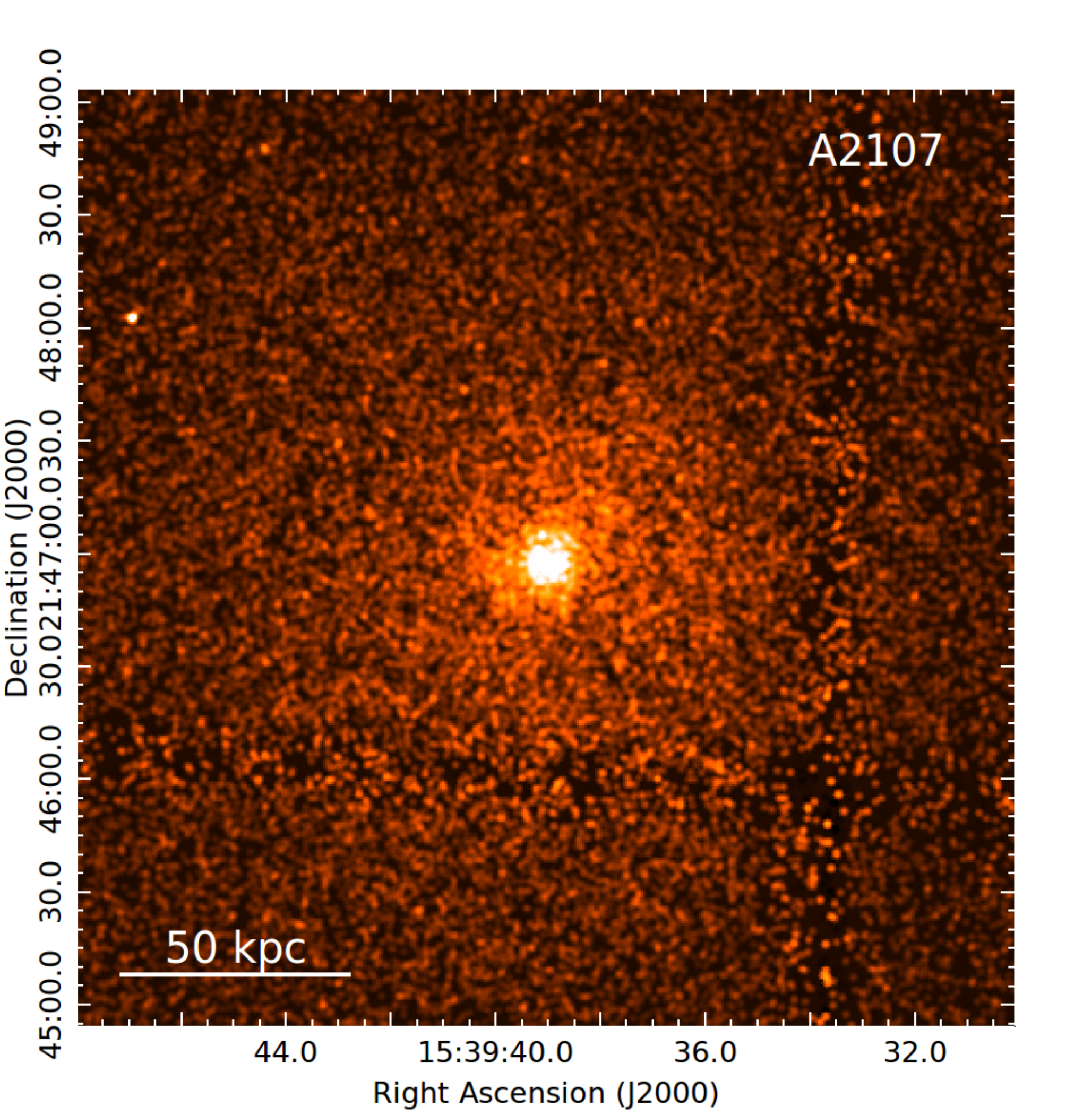}
		
		\caption{Background subtracted, and exposure corrected images of the spoiler clusters, each image is Gaussian-smoothed with a 3 arcsec kernel radius. A clearly visible negative linear feature can be seen in A2029, which is due to the absorption by a foreground spiral galaxy \citep{Clarke_2004}.}
		\label{Figure:BG_Sub_images}
		
	\end{figure*}
	
	\begin{figure*}[hp]
		
		\graphicspath{{./Profiles/plots/Images/}}
		\centering
		\includegraphics[height=76mm, width=76mm]{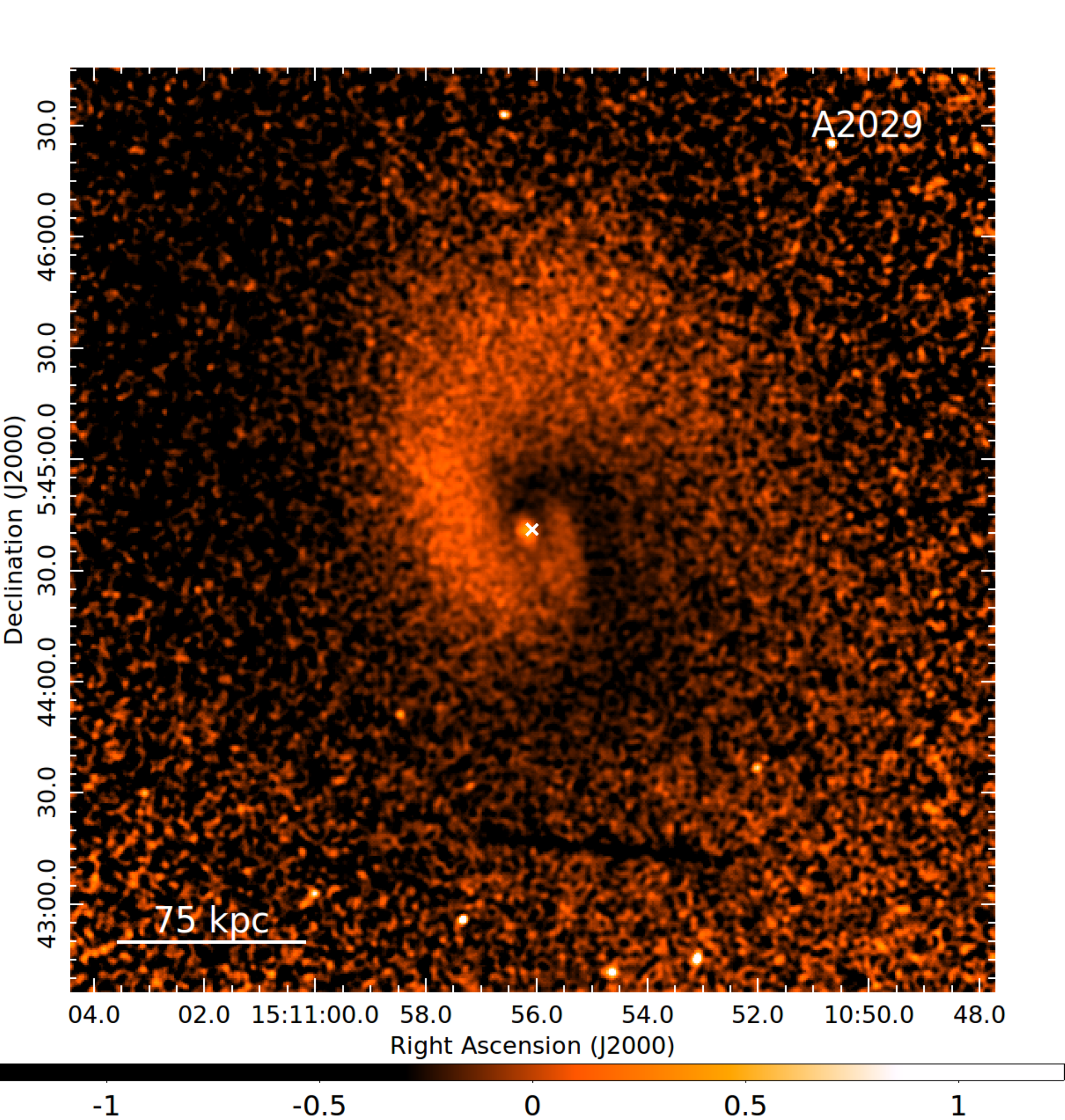}
		\includegraphics[height=76mm, width=76mm]{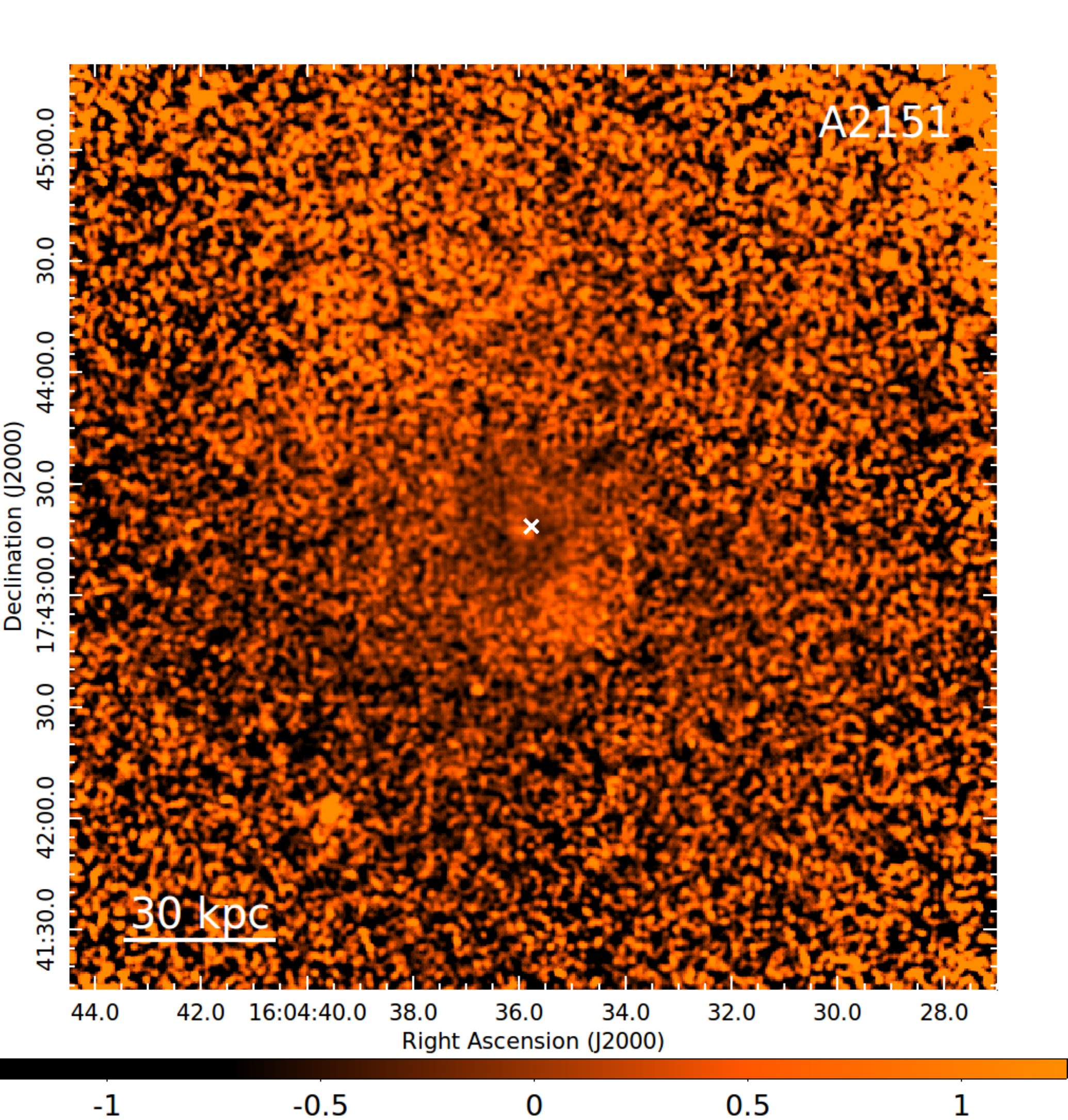}
		\includegraphics[height=76mm, width=76mm]{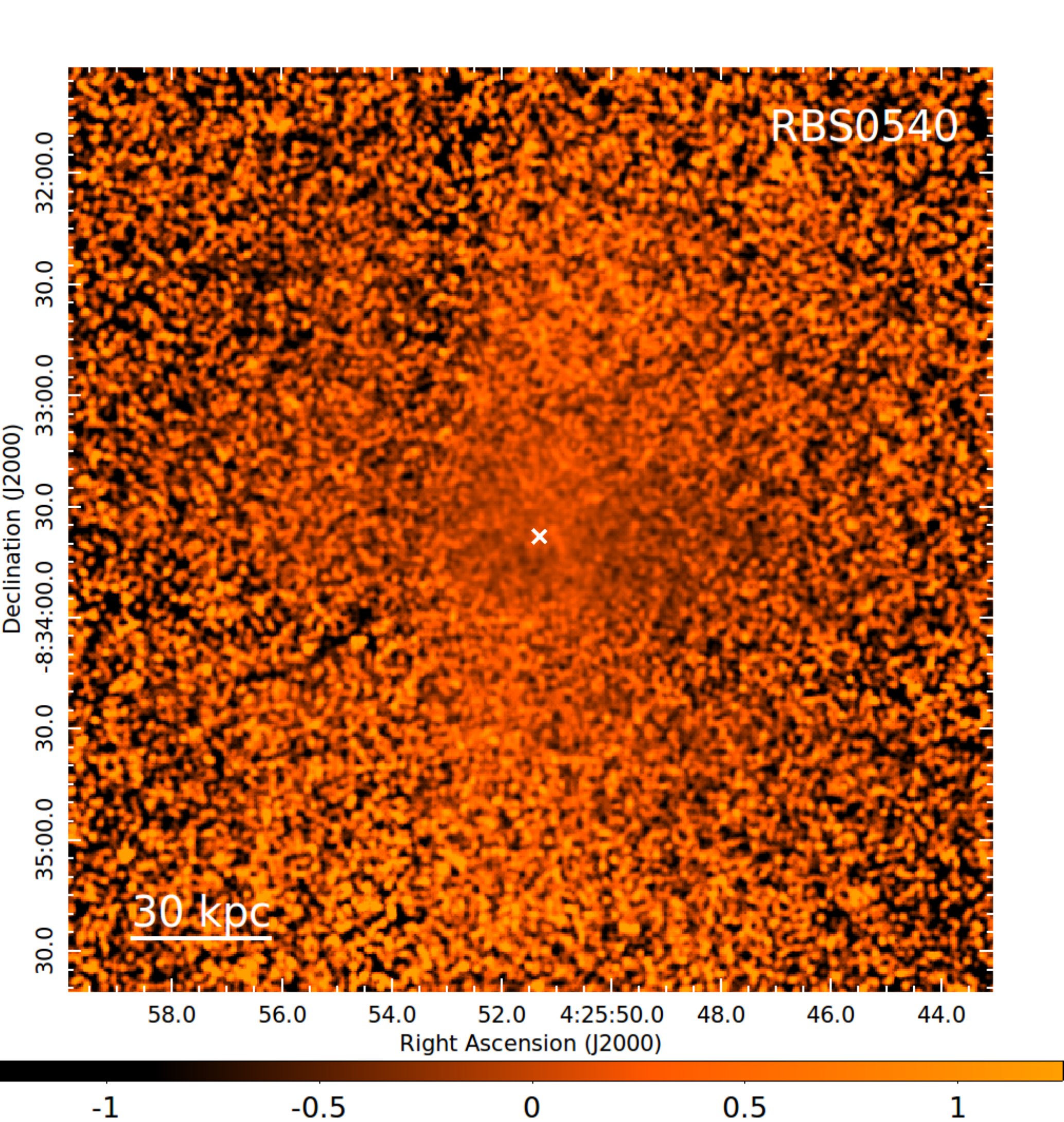}
		\includegraphics[height=76mm, width=76mm]{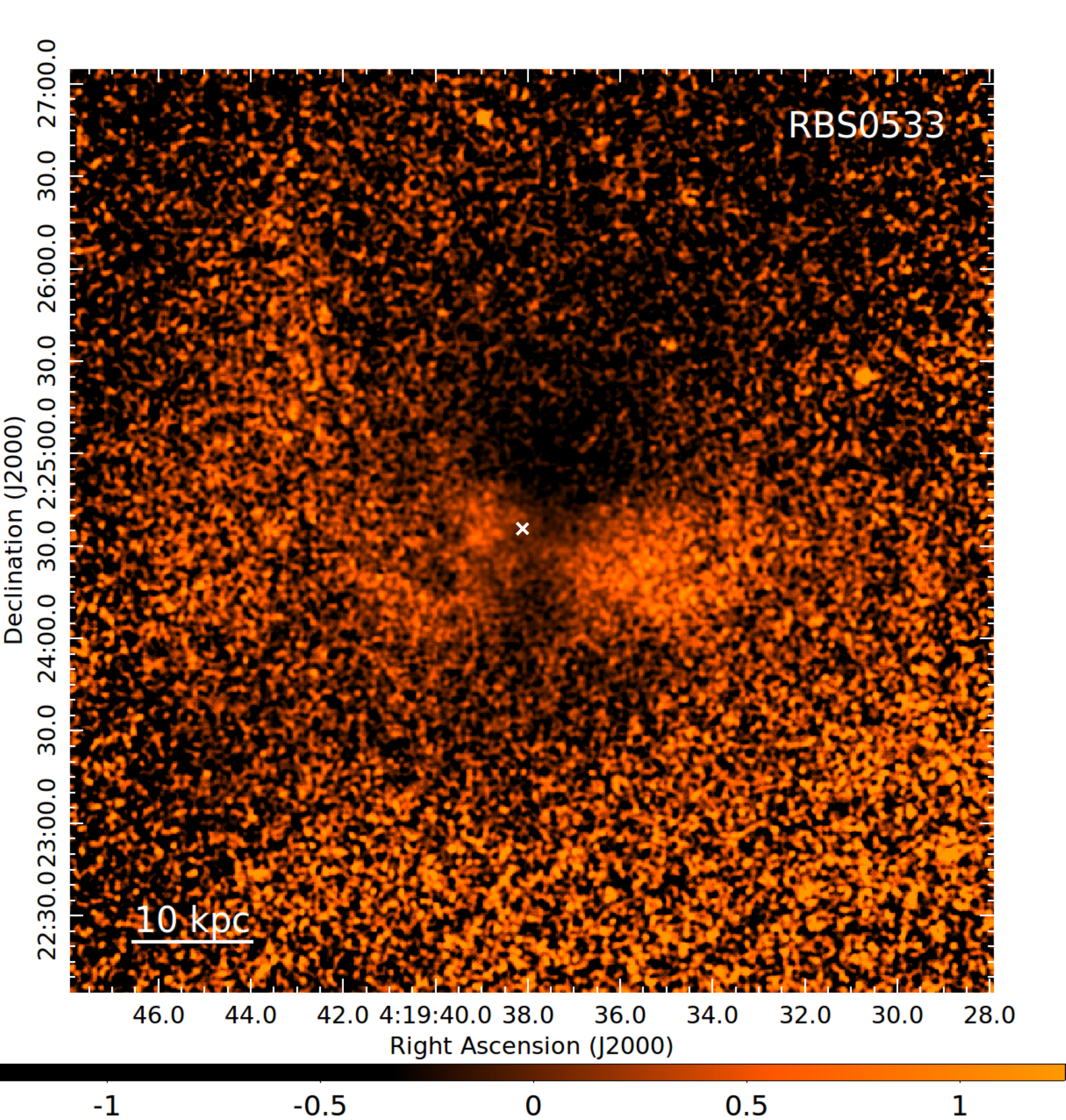}
		\includegraphics[height=76mm, width=76mm]{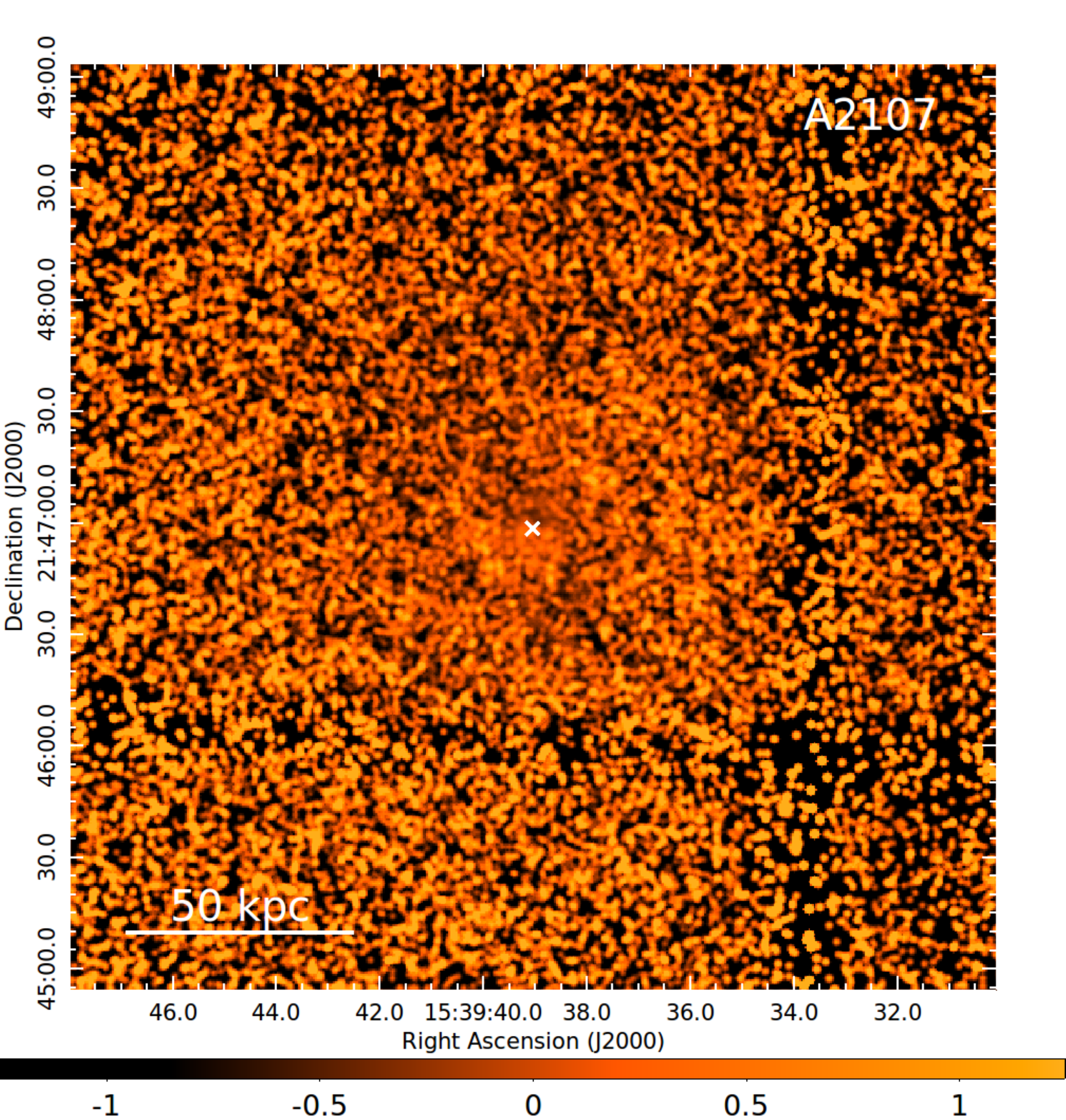}
		\caption{Model subtracted relative residual images of each cluster. Images were obtained by fitting a double $\beta$-model to the clusters SB profile of the form in Equation~\ref{Beta_Model}, and then by taking the relative difference between the images in Figure~\ref{Figure:BG_Sub_images} to that of the model. The white `x' represents the location of the brightest pixel and is the centre used in $\beta$-model fitting. These images are Gaussian-smoothed with a 3 arcsec kernel radius.}
		\label{Residual_Images}
		
	\end{figure*}

	\section{X-ray Data Analysis}
	\label{Data Analysis}
	\textit{Chandra} X-ray data were obtained for A2151 and RBS0540 and combined with pre-existing observations from the \textit{Chandra} Data Archive (CDA). The data for the remaining objects (A2029, RBS0533, A2107) were obtained from the CDA. Cluster coordinates and details of X-ray observations are shown in Table~\ref{Table:Chandra_Data}. Each observation was reprocessed using the \textsc{chandra\textunderscore repro} script with CIAO version 4.7. Bad grades were filtered out, and background light curves were extracted from level 2 event files. Events with time intervals affected by flares were eliminated using the \textsc{lc\textunderscore clean}\footnote{ http://cxc.cfa.harvard.edu/contrib/maxim/acisbg/} script.
	
	Blank-sky backgrounds for each observation were extracted using CALDB version 4.6.7. Level 2 events files (and blank-sky backgrounds) were reprojected to match the position of the observation ID (\textsc{ObsID}) with the longest exposure time. Images were constructed in the energy range of $0.5{-}7.0$ keV, for each \textsc{ObsID}. Point sources were identified using \textsc{wavdetect} \citep{Freeman_2002}, visually inspected, and removed using CIAO. Spectra were extracted from concentric circular annuli centered on the cluster's centroid. The innermost annulus was required to have a radius $\lesssim10$ kpc and binned to have a minimum of $\sim 3000$ projected counts, with the number of counts per annulus increasing within each radial bin. Spectra were extracted from these annuli separately for each \textsc{ObsID}, and were grouped with a minimum of 30 counts per energy bin.
	
	Individually weighted redistribution matrix files (RMF) and weighted auxiliary response files (ARF) were created for each spectrum using \textsc{mkacisrmf} and \textsc{mkwarf}, respectively. Exposure maps were created for each \textsc{ObsID} and used to correct for the area lost to chip gaps, point sources, and vignetting. Lastly, spectra were  deprojected using the geometric routine \textsc{dsdeproj} \citep{Sanders_2007, Russell_2008}.
	\section{Results}
	
	\subsection{Image Analysis}
	\label{Image Analysis}
	
	Evidence for disturbances in the atmospheres which could be signatures of a bubble or cavity were investigated. Surface brightness (SB) profiles of the clusters were extracted from X-ray images, for a series of concentric annuli centered on the brightest pixel. After background subtraction, the resulting surface brightness profile may be fit with an isothermal $\beta-$model \citep{Cavaliere_1976, B-R_1981}:
	\begin{equation}
	I_{X} = I_{0}[1+(R/R_{c})^{2}]^{-3\beta + 1/2},
	\label{Beta_Model}
	\end{equation}
    %
	where $I_{0}$ is the central surface brightness, $R_{c}$ is the core radius of gas distribution, and $\beta$ is the slope. The single $\beta$-model is a poor fit to the surface brightness profiles. We instead fit the surface brightness profile with a double $\beta$-model where the best fitting double $\beta$-model was subtracted from the X-ray images in Figure~\ref{Figure:BG_Sub_images}. This difference is then divided by the best fitting model to produce the residual images shown in Figure~\ref{Residual_Images}. This procedure accentuates fluctuations in the surface brightness of the ICM revealing substructure. The residual image also reveals structure to the North of RBS0533 which may be a cavity or bubble. In Section~\ref{SB Variations}, we estimate the significance of this region and draw conclusions on whether or not this structure is a bubble. The remaining clusters show evidence of variations in surface brightness, although none of these are likely due to bubbles or AGN feedback.
	
    For instance, sloshing of intracluster gas can create sharp changes in temperature and density that appear as surface brightness edges when projected onto the sky \citep{Markevitch&Vikhlinin_2007}. These cold fronts are created by merging halos which displace low entropy gas from the centre of the potential which wraps into a spiral feature due to sloshing motions. Such spiral features have been observed in many clusters, such as Perseus \citep{Fabian_2006}, Virgo \citep{Roediger_2011}, Centaurus \citep{Sanders_2016}, A496 \citep{Roediger_2012}, and A2029 \citep{Paterno-Mahler_2013}. 
	
	\cite{Roediger_2012} performed hydrodynamic simulations to explore the nature and origin of sloshing spiral features in A496. Cold fronts created in this simulation combined to form the observed spiral features, which are not necessarily a result of recent mergers. \cite{Ascasibar&Markevitch_2006} have shown that such features can persist for several Gyrs. We explored our clusters for evidence of such features in the ICM.

	In A2029, a cold front is clearly visible as a sharp change in surface brightness as seen in the top left image of Figure~\ref{Figure:BG_Sub_images}. Our residual image of A2029, shown in Figure~\ref{Residual_Images}, confirms the continuous spiral feature directly associated with the cold front. Its spiral feature is the largest and most continuous one known, extending outward radially from the centre up to approximately 400 kpc \citep{Paterno-Mahler_2013}.
	
	Similarly, A2151's residual image also reveals a possible spiral feature. While not as prominent as A2029's sloshing feature, it extends radially outwards to at least 81 kpc. We find no clear evidence of sloshing in the remaining objects.
	
	\subsection{Projected Thermodynamic Profiles}
	\label{Projected}
	Spectra for each annular region were obtained using the method described in Section~\ref{Data Analysis}. Spectra were fit with the absorbed thermal model, \textsc{phabs(apec)}. Abundances were determined relative to the values of \cite{Anders_1989}. Excluding the redshift (frozen to the value of the cluster), all parameters -- the column density of neutral hydrogen ($\rm N_{ H}$), temperature, metallicity, and normalization -- were allowed to vary. These values were used to derive the projected electron densities ($n_{e}$),
	\begin{equation}
	n_{e}=D_{A}(1+z)10^{7}\sqrt{\dfrac{1.2 \ N\ 4\pi}{V}},
	\end{equation}
	%
	where $z$ is the redshift, $N$ is the model normalization, $D_{A}$ is the angular diameter distance to the source, and $V$ is the volume of a spherical shell with the inner and outer radii set to the annulus edges. The 1.2 factor is the ratio of the electron density to the hydrogen number density, $n_{\rm e}/n_{\rm H}$ \citep{Anders_1989}. The cooling time of the ICM was calculated as
	\begin{equation}
	t_{\rm cool}=\dfrac{3p}{2n_{e}n_{H}\Lambda(Z,T)}=\dfrac{3pV}{2L_{X}},
	\end{equation}
	%
	where $p$ is the pressure, $p=1.8n_{e}kT$, $L_{X}$ is the X-ray luminosity within each shell, and $\Lambda(Z,T)$ is the gas cooling function in terms of abundance and temperature. $L_{X}$ was determined by first refitting spectra with a \textsc{phabs$\times$cflux(apec)} model. We obtained an estimate for the X-ray bolometric flux by integrating the unabsorbed thermal model between $0.1$ and $100$ keV. The resulting X-ray flux was then converted to a bolometric X-ray luminosity, $L_{X}$. 
	
\begin{figure*}[!ht]
		
		\graphicspath{{./profiles/plots/projected_properties/}}
		\centering
		
		\includegraphics[height=64mm, keepaspectratio]{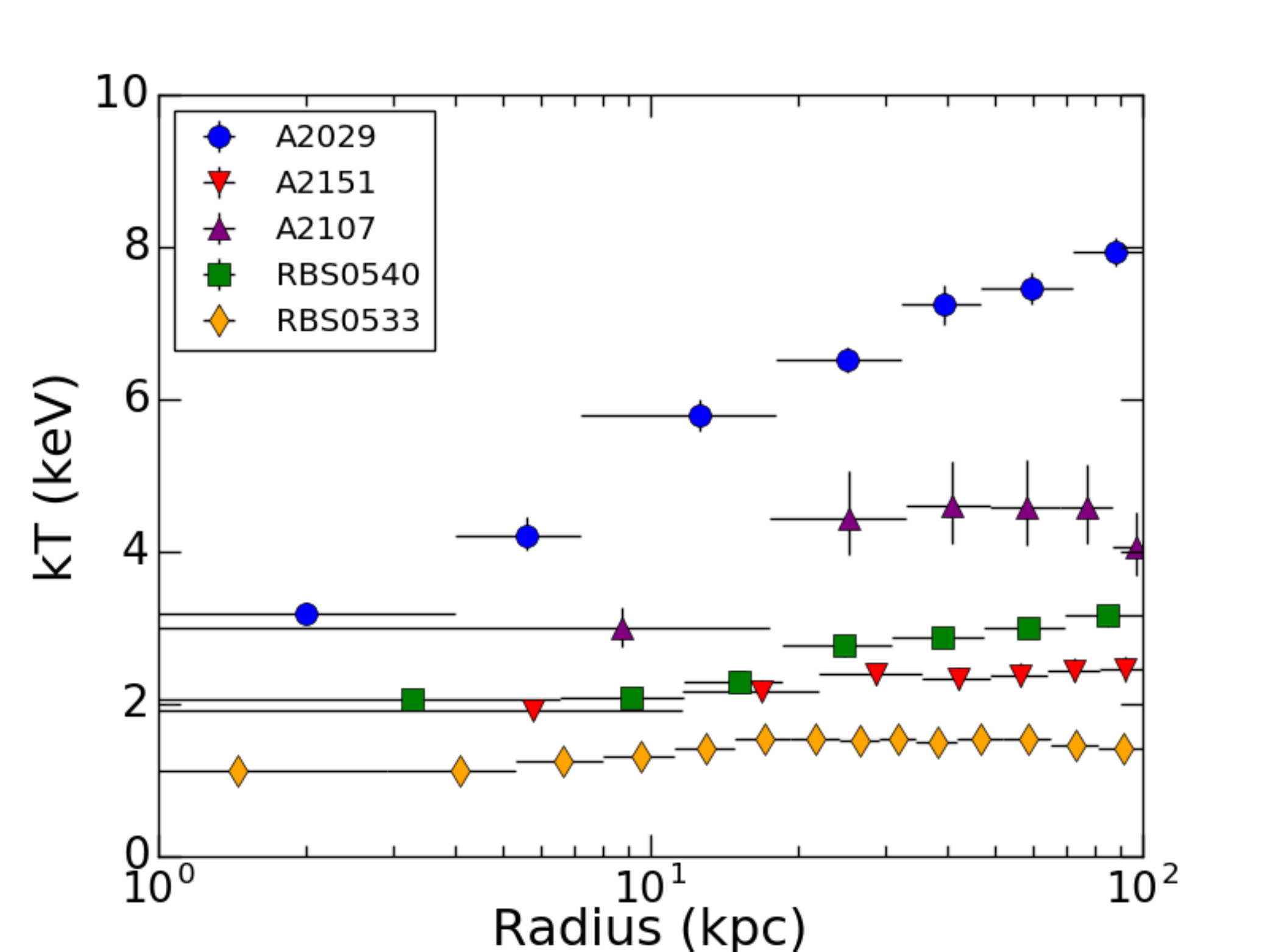}
		\includegraphics[height=64mm,keepaspectratio]{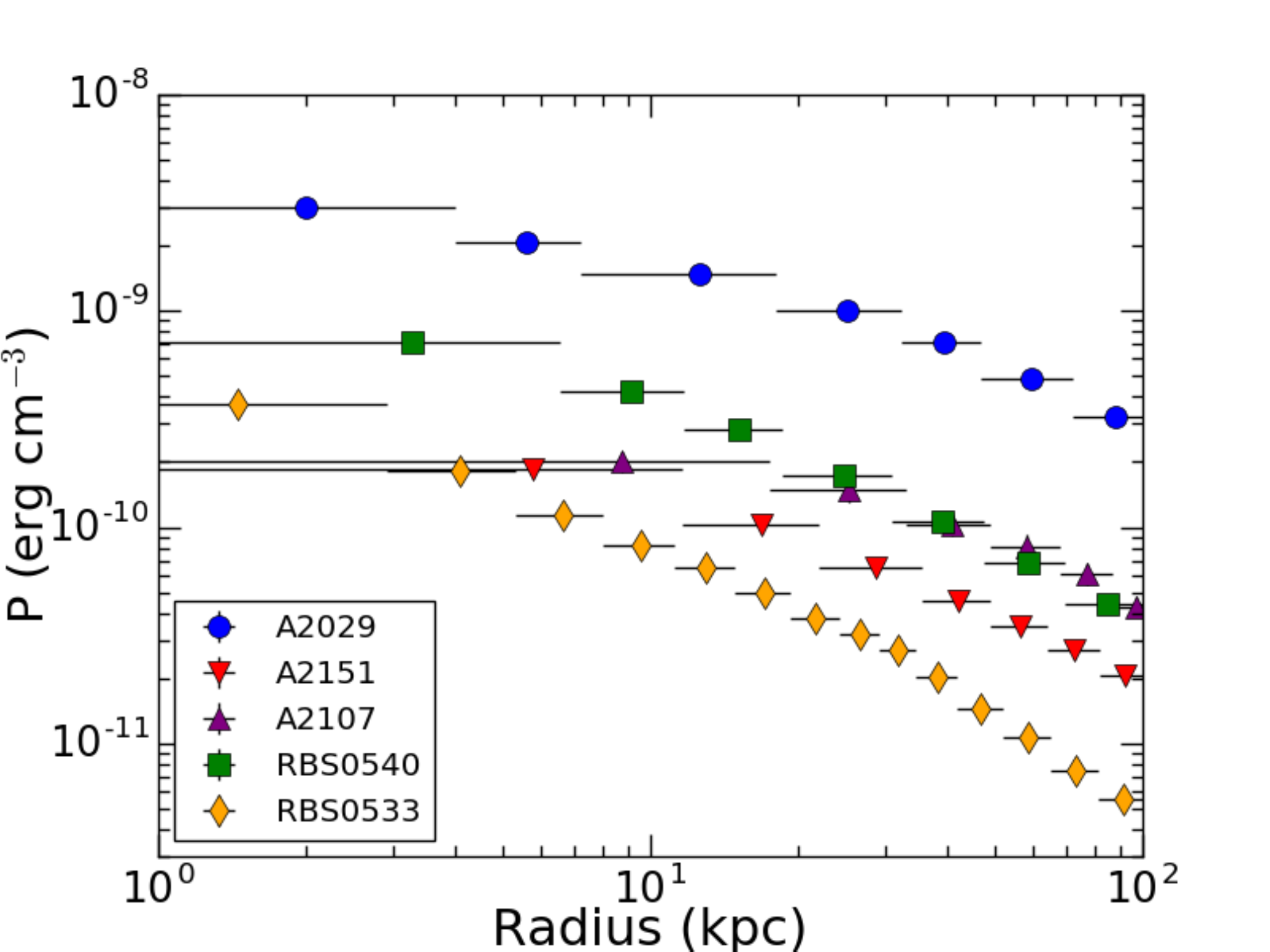}
		\includegraphics[height=64mm,keepaspectratio]{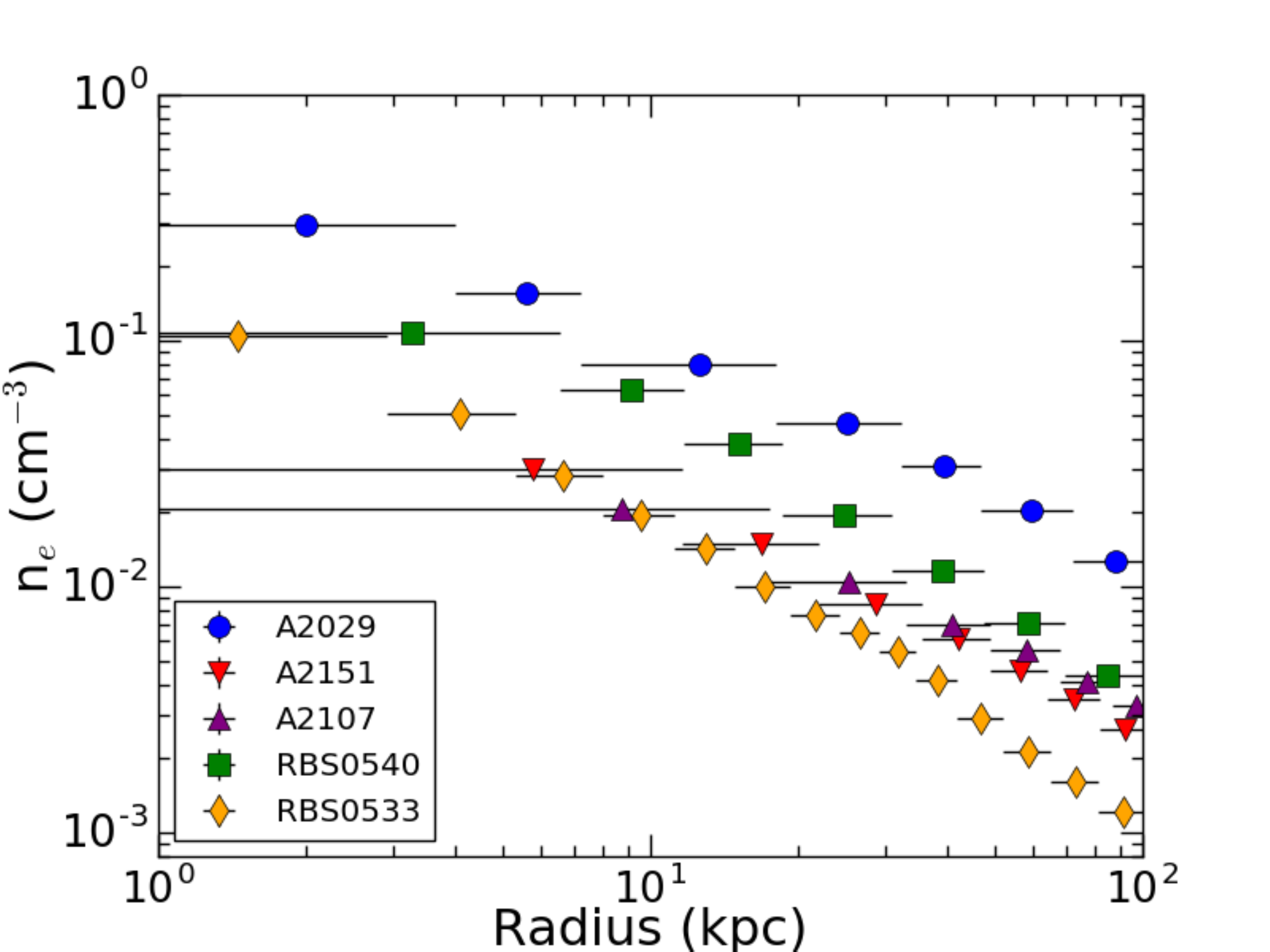}
		\includegraphics[height=64mm,keepaspectratio]{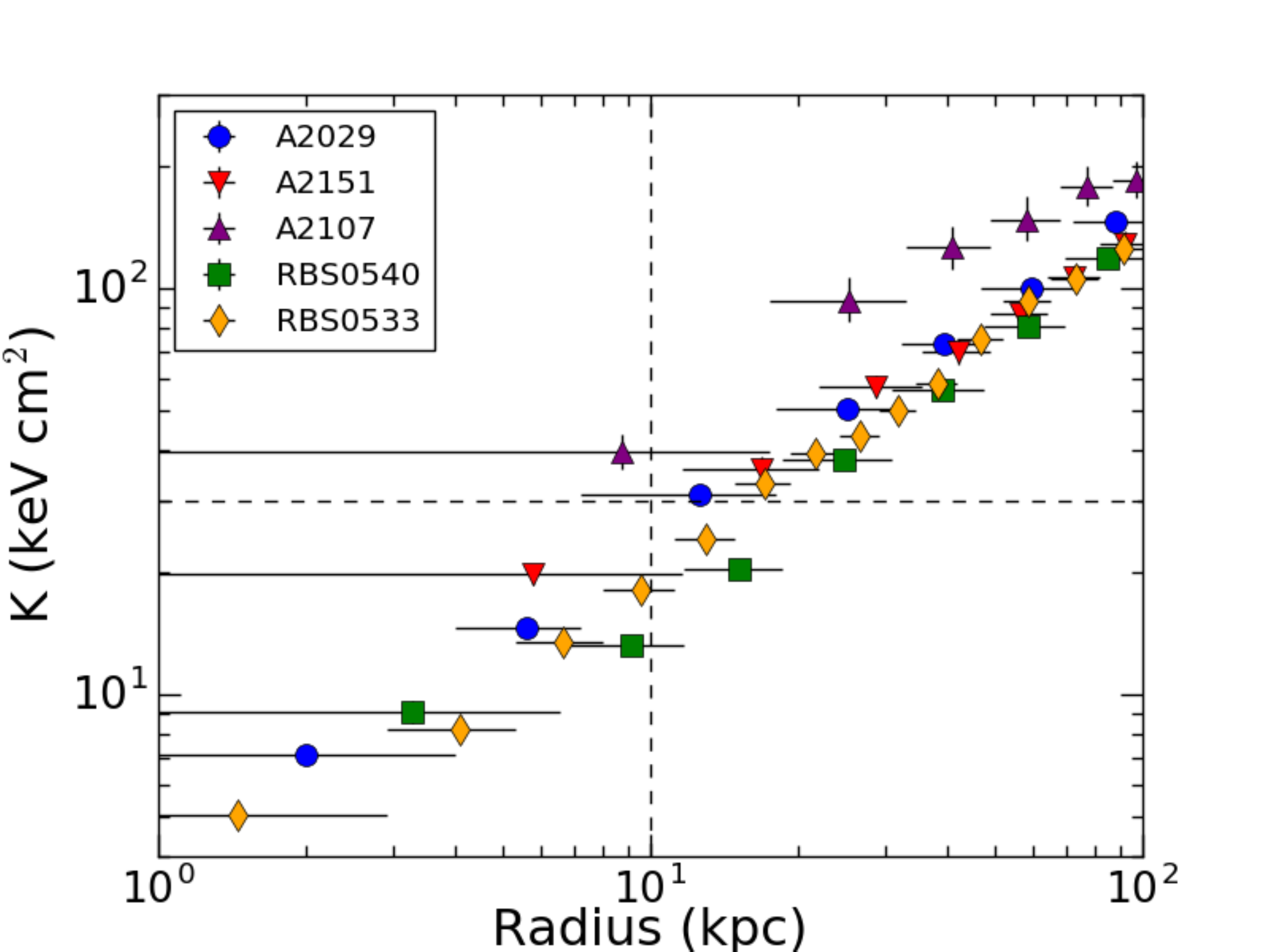}
		\includegraphics[height=64mm,keepaspectratio]{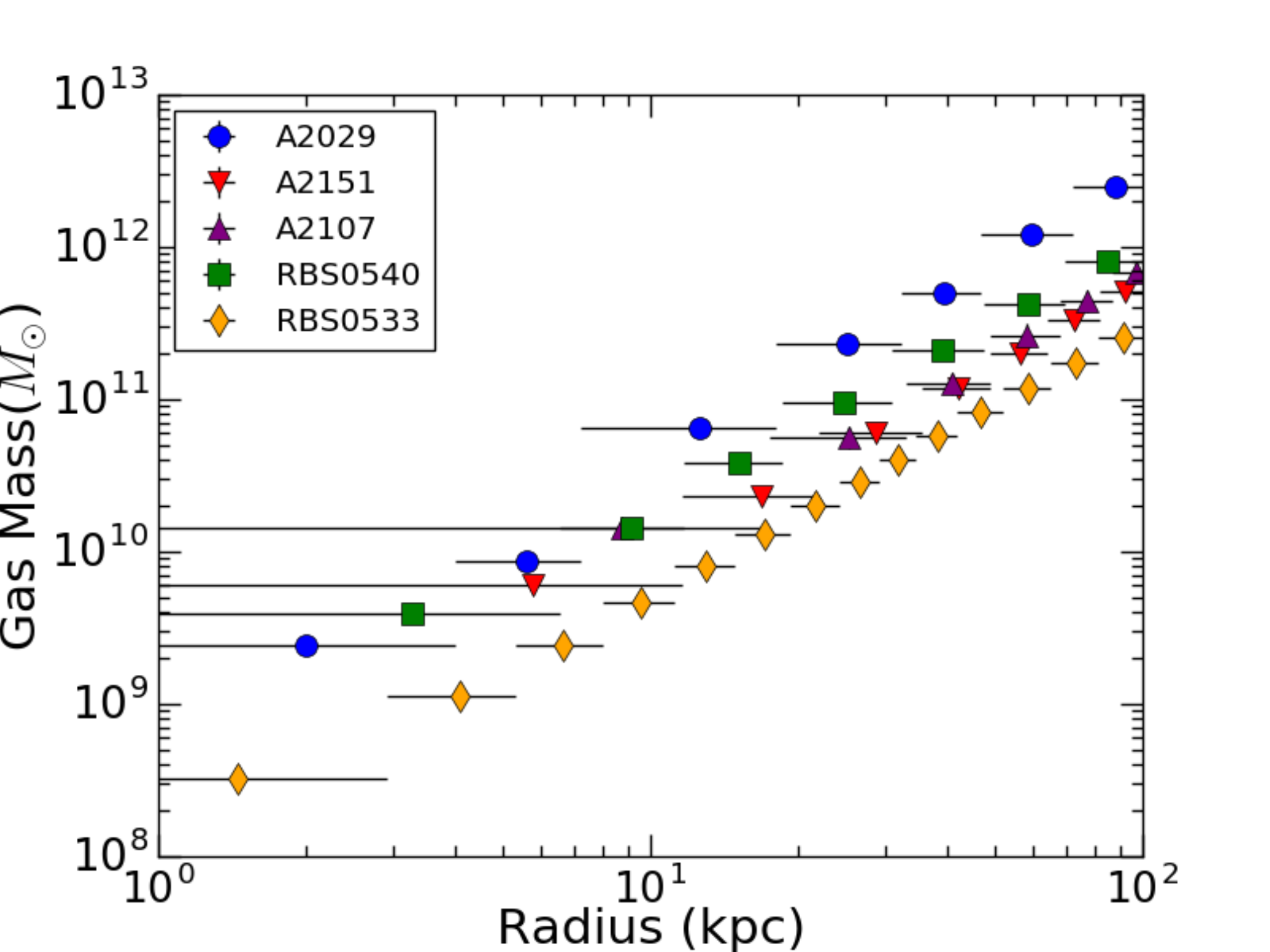}
		\includegraphics[height=64mm,keepaspectratio]{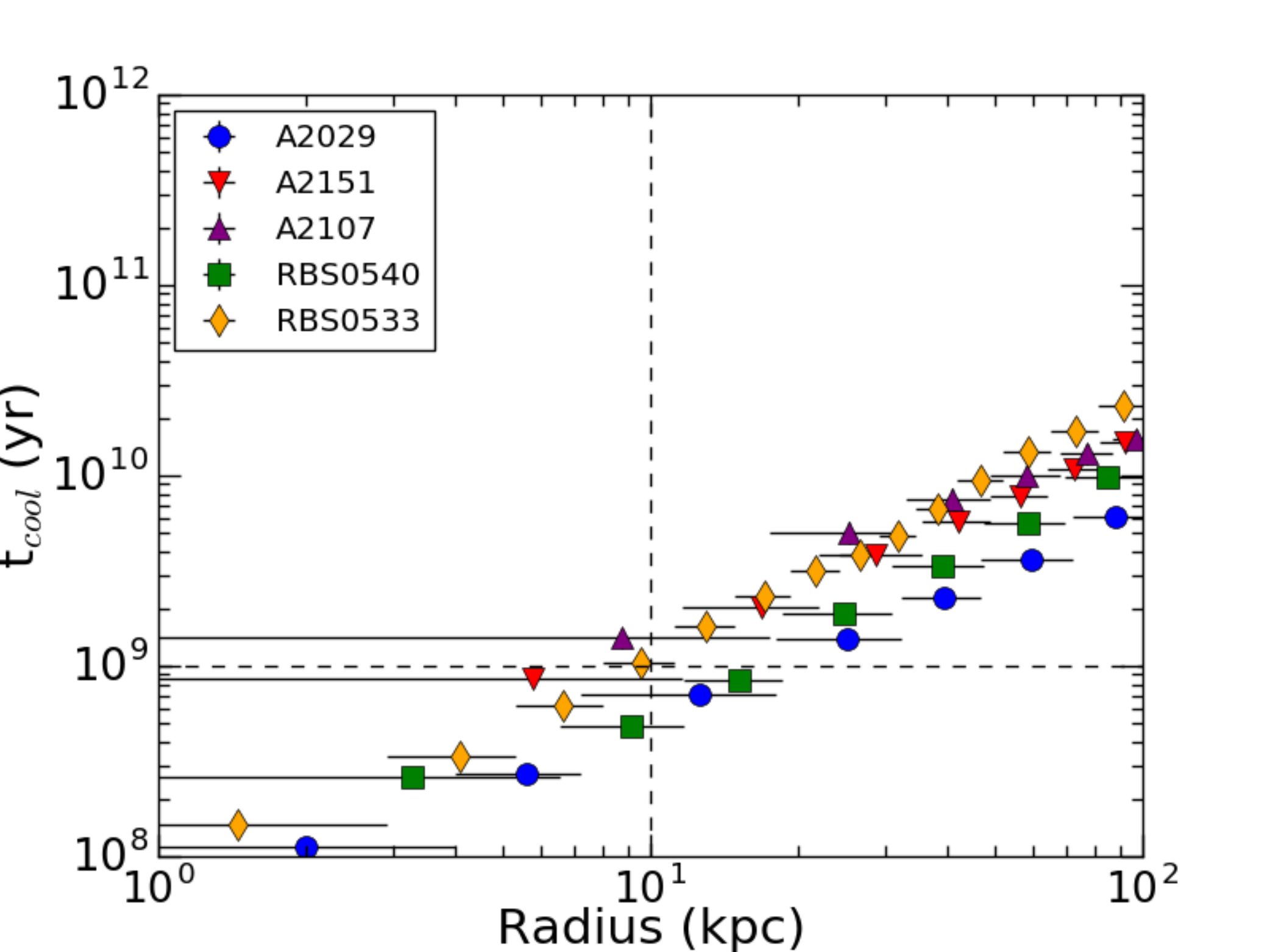}
		
		\caption{Projected  temperature, pressure, density, entropy, gas mass, and cooling time profiles. The dashed horizontal lines represent the thresholds for the cooling time ($t_{cool}=1.0 \times 10^{9}$ yr) and entropy (K$=30$ keV cm$^{2}$). All errors here are reported at the $2\sigma$ level.}
		\label{Figure:Projected_Profiles}
		
	\end{figure*}

	Finally, we derived the entropy parameter, $K=kT n_{e}^{-2/3}$, and hot gas mass within each spherical shell, $M=1.9\mu m_{p} n_{e} V$, where $m_{p}$ is the proton mass, and $\mu= 0.62$ represents the mean molecular weight of atmospheric plasma. Projected profiles are shown in Figure~\ref{Figure:Projected_Profiles}. Of these, A2029 spans a substantially larger temperature ($3.0-9.5$ keV) than the other clusters. Its density is nearly an order of magnitude higher than the other clusters, however it still satisfies the threshold for central cooling time in projection.

	\subsection{Deprojected Thermodynamic Profiles}
	\label{Deprojected}
	
\begin{figure*}[!ht]
		
		\graphicspath{{./profiles/plots/deprojected_properties/}}
		\centering
		\includegraphics[height=64mm, keepaspectratio]{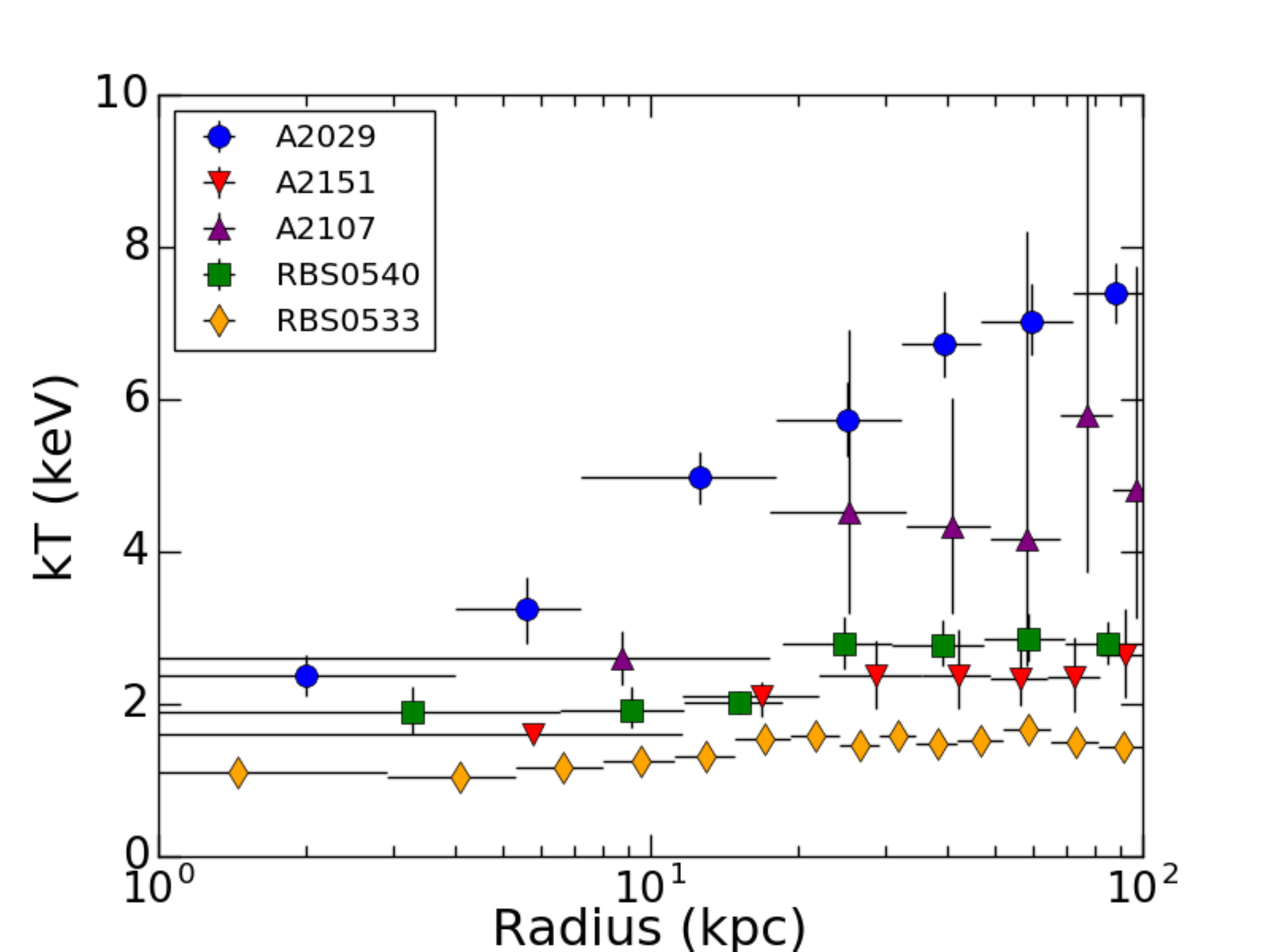}
		\includegraphics[height=64mm, keepaspectratio]{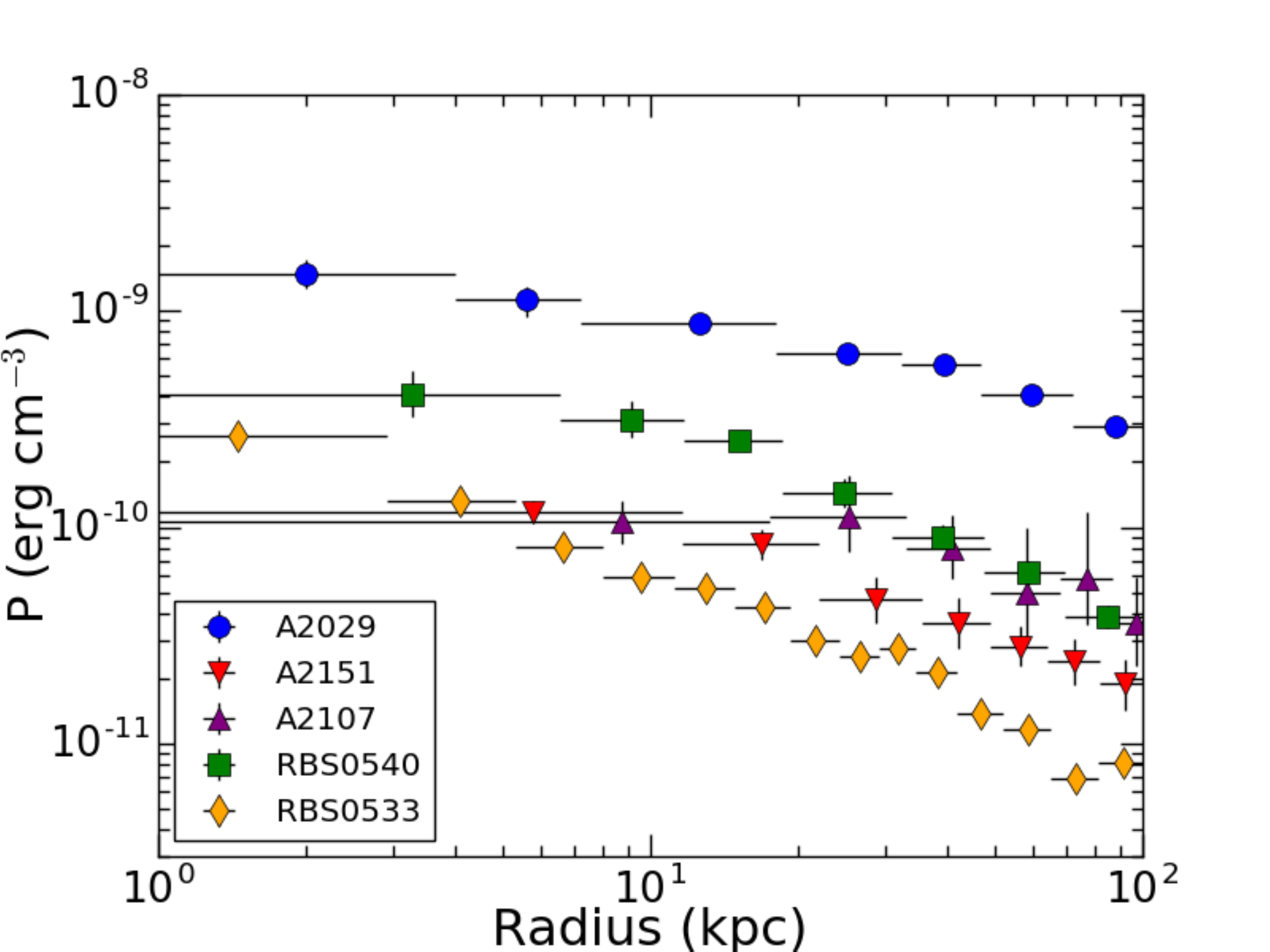}
		\includegraphics[height=64mm, keepaspectratio]{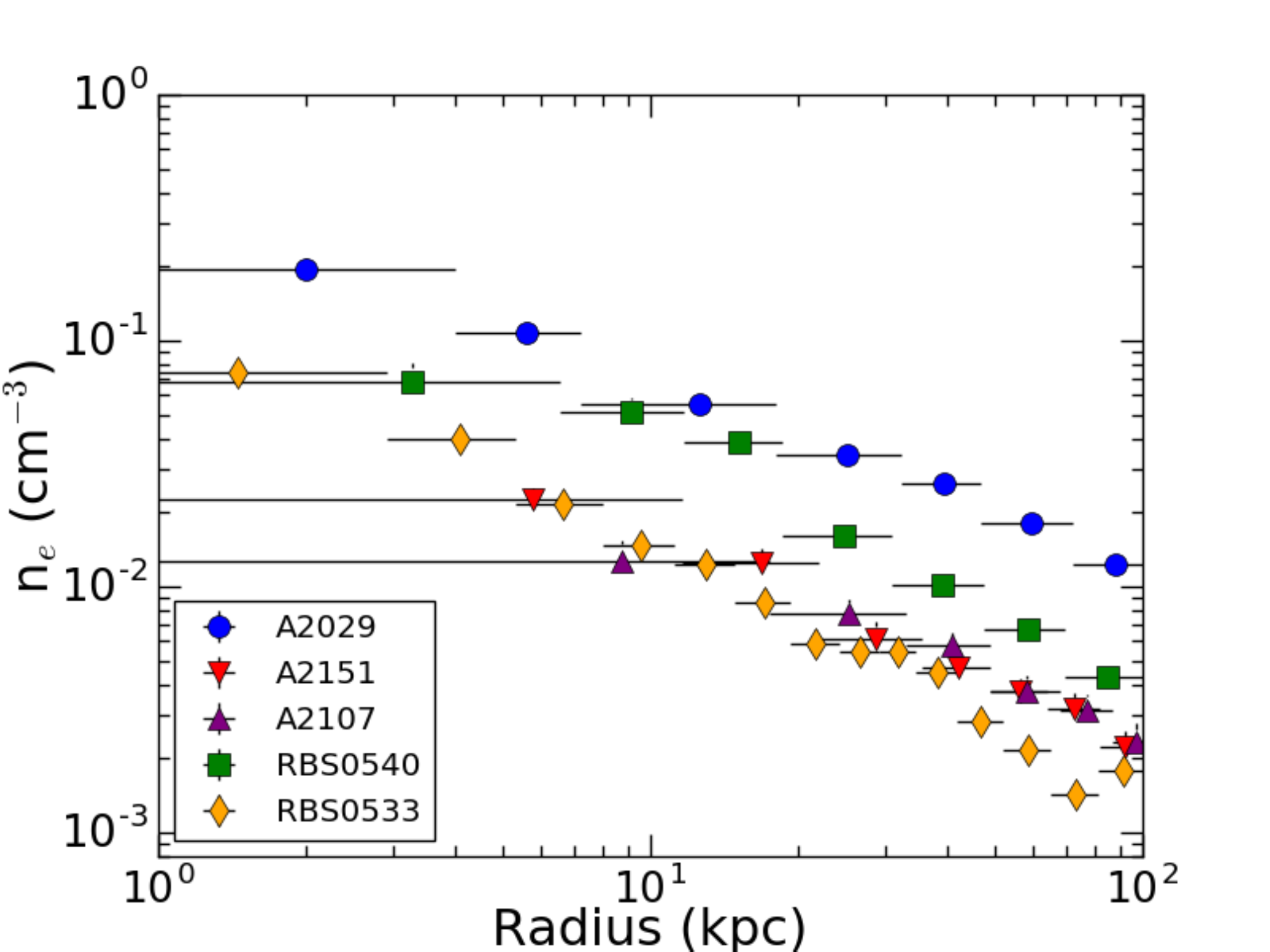}
		\includegraphics[height=64mm, keepaspectratio]{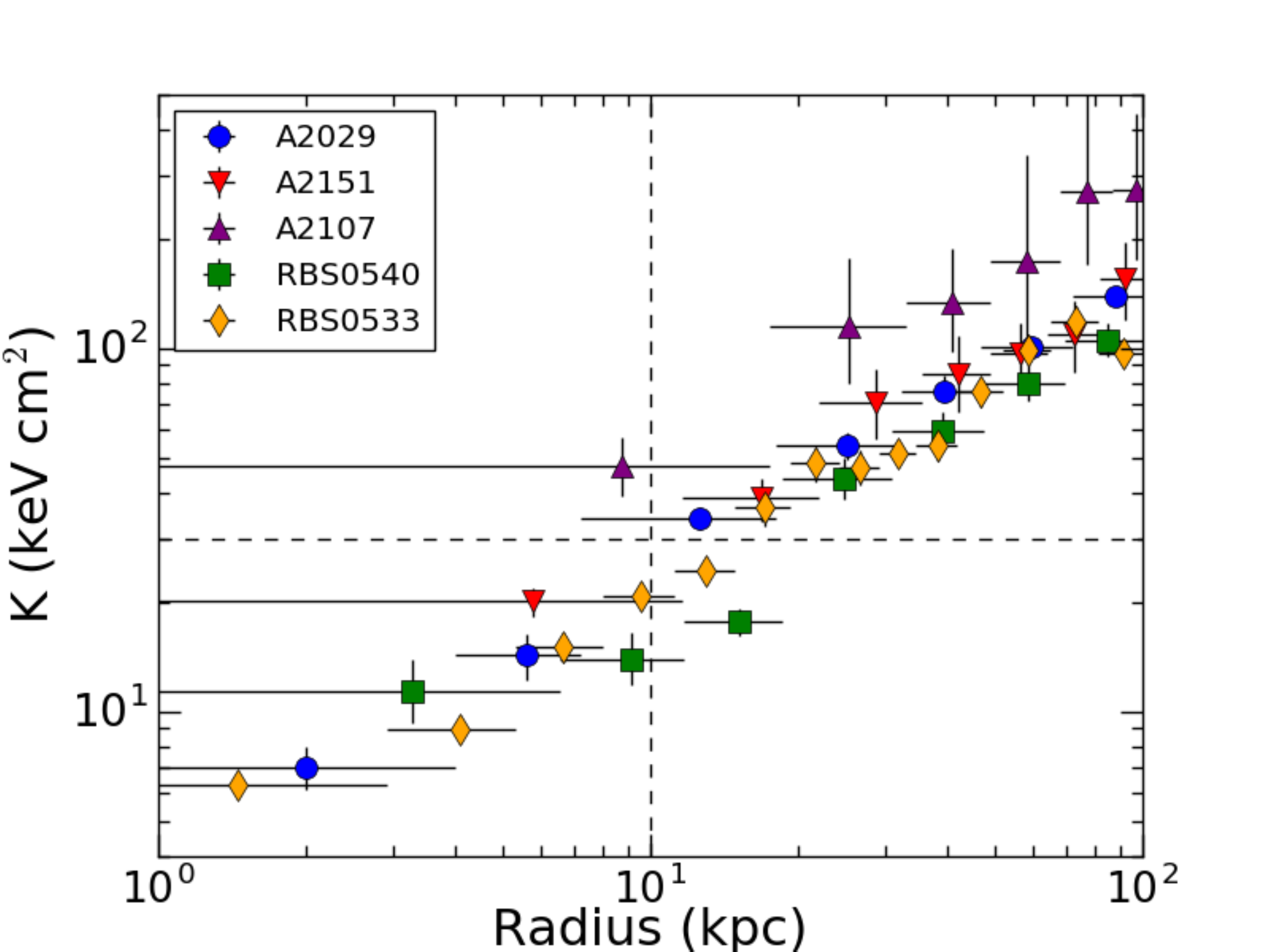}
		\includegraphics[height=64mm, keepaspectratio]{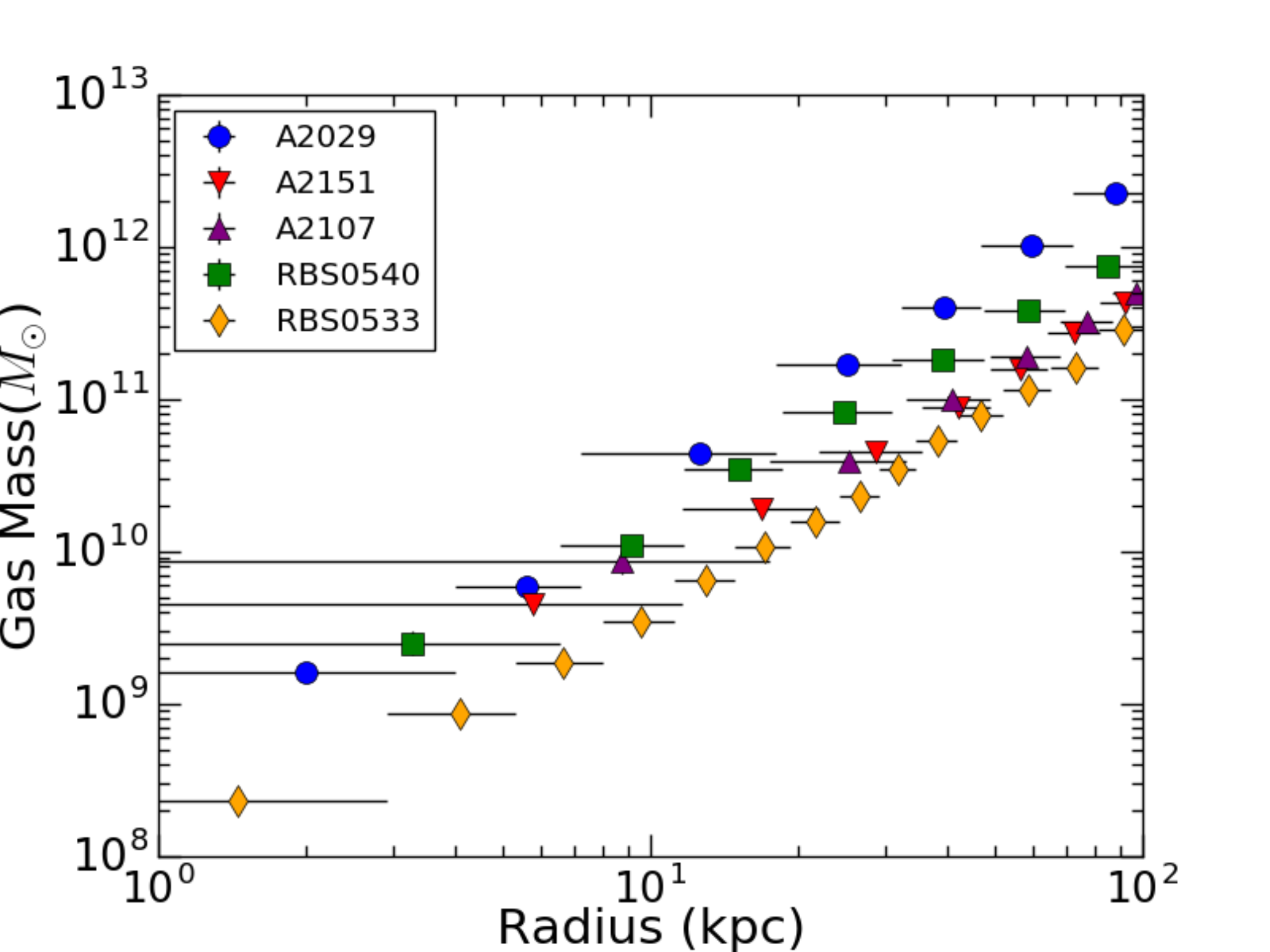}
		\includegraphics[height=64mm, keepaspectratio]{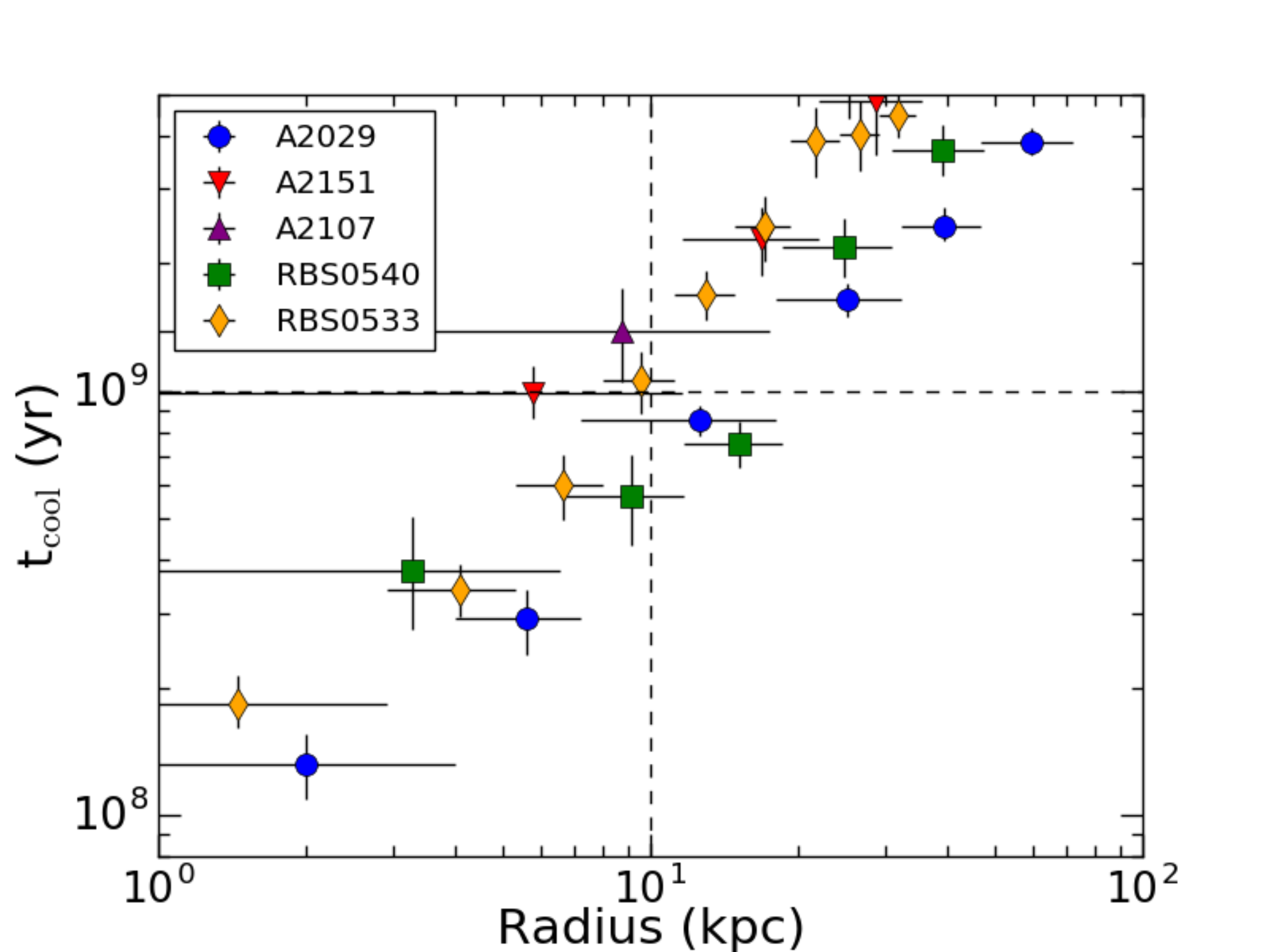}
		
		\caption{Deprojected temperature, pressure, density, entropy, gas mass, and cooling time profiles. The dashed horizontal lines represent the thresholds for the cooling time ($t_{cool}=1.0 \times 10^{9}$ yr) and entropy (K$=30$ keV cm$^{2}$). All errors here are reported at the $2\sigma$ level.}
		\label{Figure:Deprojected_Profiles}
		
	\end{figure*}

	Spectra extracted from the inner regions are affected by projected emission from the hotter regions of the atmosphere at higher altitudes. Accurately deriving the profiles of the inner regions of a cluster requires deprojection. This was done using the \textsc{dsdeproj} routine \citep{Russell_2008}. Similar to our projected profiles, we fit the deprojected spectra to a single-temperature \textsc{phabs(apec)} model. Again, all parameters except for redshift were allowed to vary. Fitted quantities for temperature, abundance, and model normalization were used to derive the deprojected density, cooling time, entropy, and the hot gas mass profile, as shown in Figure~\ref{Figure:Deprojected_Profiles}. The deprojected profiles indicate that within the central region of each source ($\lesssim$10 kpc), both the cooling time and the entropy lie below 1 Gyr and 30 keV cm$^{2}$, respectively. 
	
	These objects span a moderate range of temperatures between $1-8$ keV, while their radial densities and pressures span at least 2 orders of magnitude from $10^{-1}-10^{-3}$ cm$^{-3}$ and $10^{-9}-10^{-11}$ erg cm$^{-3}$, respectively.
	
	\subsection{Mass Profiles}
    \label{Cluster Masses}
	
	Mass profiles were created following the model presented in \cite{Hogan_2017a}. The model consists of a Navarro-Frenk-White (NFW) potential and a cored isothermal potential. The former has been shown to be an accurate description of the total gravitating potential of cluster masses on large scales  \cite[e.g.][]{Pointecouteau_2005,Vikhlinin_2006} and takes the form of
	\begin{equation}
	\Phi_{\rm NFW}(R) = -4\pi G \rho R_{s}^{2} \frac{\ln\big(1 + R/R_{s}\big)}{R/R_{s}},
	\end{equation}
	%
	where $\rho$ is the characteristic gas density and $R_{s}$ is the scale radius. Although the NFW profile provides a reasonable fit on large scales, its contribution alone underestimates masses inferred from stellar velocity dispersion due to the central galaxy \cite[e.g.][]{fisher_1995, Lauer_2014, Hogan_2017a}. Thus, a cored isothermal potential is needed to account for this, given by
	\begin{equation}
	\Phi_{\rm ISO}(R) = \sigma_{*}^{2}\ln\big(1 + (R/R_{I})^{2}\big).
	\end{equation}
	\indent Here $\sigma_{*}$ is the stellar velocity dispersion and $R_{I}$ which is used solely to prevent the gravitational acceleration from diverging from $R \to 0$, is chosen to be smaller than the scales of interest. The combined NFW and cored isothermal potential, \textsc{isonfwmass}, is implemented as an \textsc{xspec} extension in the package \textsc{clmass} \citep{Nulsen_2010}. The model assumes the cluster atmosphere is spherically symmetric and in hydrostatic equilibrium. 
	
	Stable fits were found by following the work of \cite{Hogan_2017a}. The $\sigma_{*}$ parameter was frozen to the inferred stellar velocity dispersions derived from 2MASS isophotal K-band magnitude. When unavailable, values were taken from the HyperLEDA database \citep{HyperLeda_2014}, or assumed to be 250 km s$^{-1}$ when no data was available \citep{Voit_Donahue_2015}.

    \begin{table*}[ht]
	\begin{center}
		\caption{Mass Fitting Parameters.}
		\label{Mass_Profile_Fits}
		\scalebox{1.25}{
		\begin{threeparttable}[whiteb]
			
			\begin{tabular}{@{}lcccccc}
				\hline\hline
				Cluster & $\sigma_{*}$ & $A_{\rm{ISO}}$ & $R_{s}$ & $A_{\rm{NFW}}$ & $R_{2500}$ & $M_{2500}$  \\
				& (km s$^{-1}$) & (keV) & (arcmin) & (keV) & (kpc) & ($10^{14}$ M$_{\odot}$) \\
				& (1) & (2) & (3) & (4) & (5) & (6) \\
				\Xhline{1.5\arrayrulewidth}
				\\
                A2029 & $336\pm10^{a}$ & $0.694$ & $5.35^{+0.39}_{-0.28}$ & $86.5^{+4.6}_{-4.2}$ & $693.4$ & $5.1^{+0.20}_{-0.18}$ \\[5pt]
                \Xhline{2.5\arrayrulewidth}
                \\
	            A2151 & $219\pm4^{a}$ & $0.295$ & $1.87^{+0.29}_{-0.13}$ & $9.7^{+0.5}_{-0.4}$ & $259.2$ & $0.26^{+0.01}_{-0.01}$ \\[5pt]
	            \Xhline{2.5\arrayrulewidth}
                \\
	            A2107 & $314\pm 25^{b}$ & $0.608$ & $6.32^{+1.80}_{-0.88}$ & $26.8^{+7.1}_{-3.5}$ & $414.6$ & $1.05^{+0.22}_{-0.23}$ \\[5pt]
	            \Xhline{2.5\arrayrulewidth}
                \\
	            RBS0533 & $306\pm14^{b}$ & $0.575$ & $13.8^{+9.0}_{-5.4}$ & $4.3^{+2.3}_{-1.4}$ & $228.3$ & $0.17^{+0.02}_{-0.02}$\\[5pt]
	            \Xhline{2.5\arrayrulewidth}
                \\
	            RBS0540 & $250\pm15^{c}$ & $0.384$ & $1.17^{+0.24}_{-0.17}$ & $12^{+0.91}_{-0.72}$ & $279.5$ & $0.32^{+0.03}_{-0.03}$ \\[5pt]
				\hline\hline
			\end{tabular}
			\begin{flushleft}\textbf{Note.} Columns are: (1) Equivalent stellar velocity dispersion, (2) Isothermal potential given by, $A_{\rm{ISO}} = \mu_{0} m_{p}\sigma_{*}^{2}$, (3) NFW scale radius, (4) NFW potential given by,  $A_{\rm{NFW}}= 4 \pi \mu_{0} m_{p} G \rho R_{s}^{2}$, (5) $R_{2500}$, (6) $M_{2500}$. 
	        \\ $^{\textbf{a}}$ $\sigma_{*}$ inferred from 2MASS isophotal K-band magnitude measurements.
	        \\ $^{\textbf{b}}$ $\sigma_{*}$ measurements from HyperLEDA.
	        \\ $^{\textbf{c}}$ Assuming $\sigma_{*}=250$ km s$^{-1}$, following \cite{Voit_Donahue_2015}.\end{flushleft}
		\end{threeparttable}}
	\end{center}		
    \end{table*}
  
	\begin{figure*}[ht]
		\centering
		\includegraphics[width=0.46\linewidth, keepaspectratio]{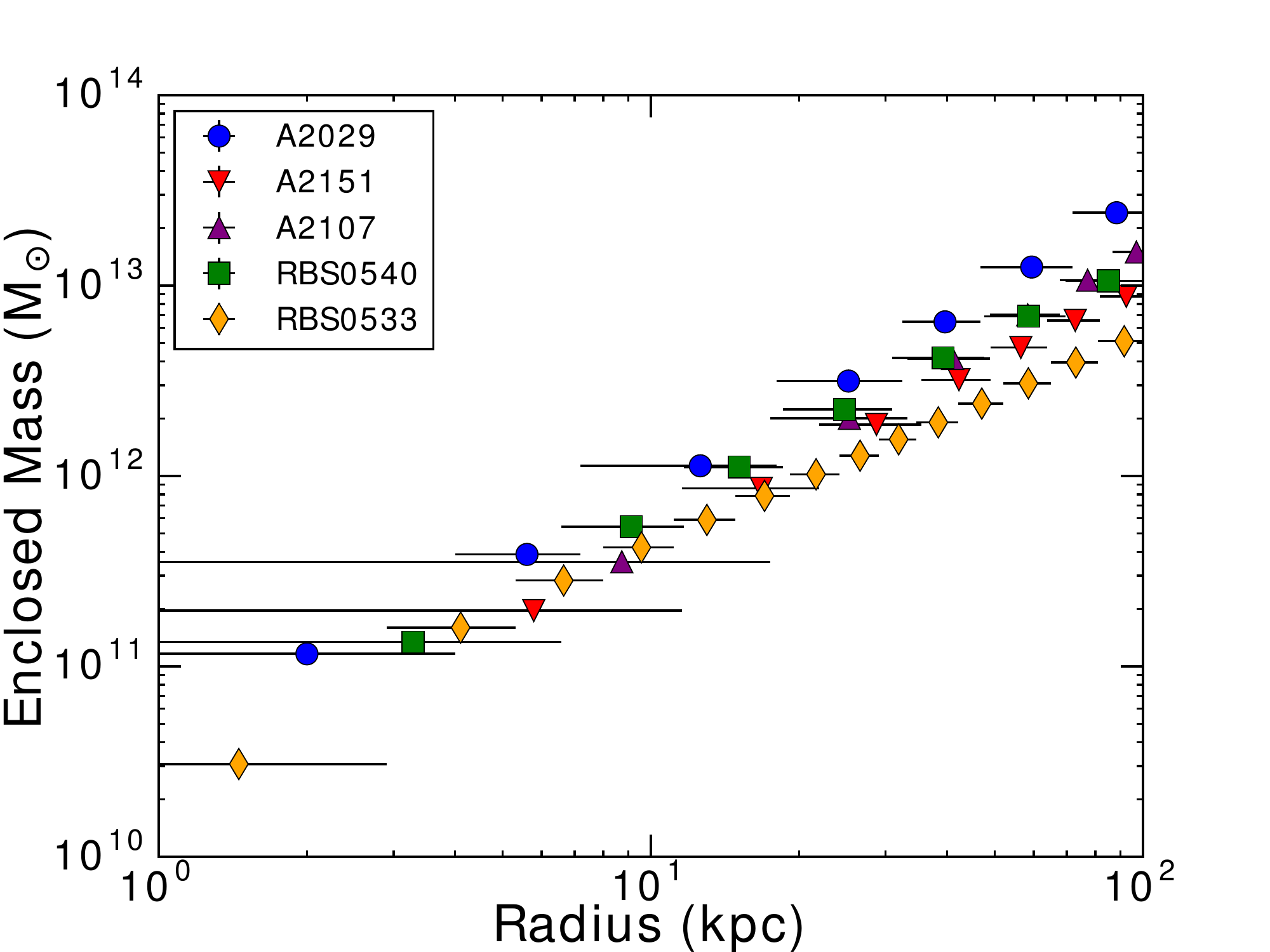}
		\includegraphics[width=0.46\linewidth, keepaspectratio]{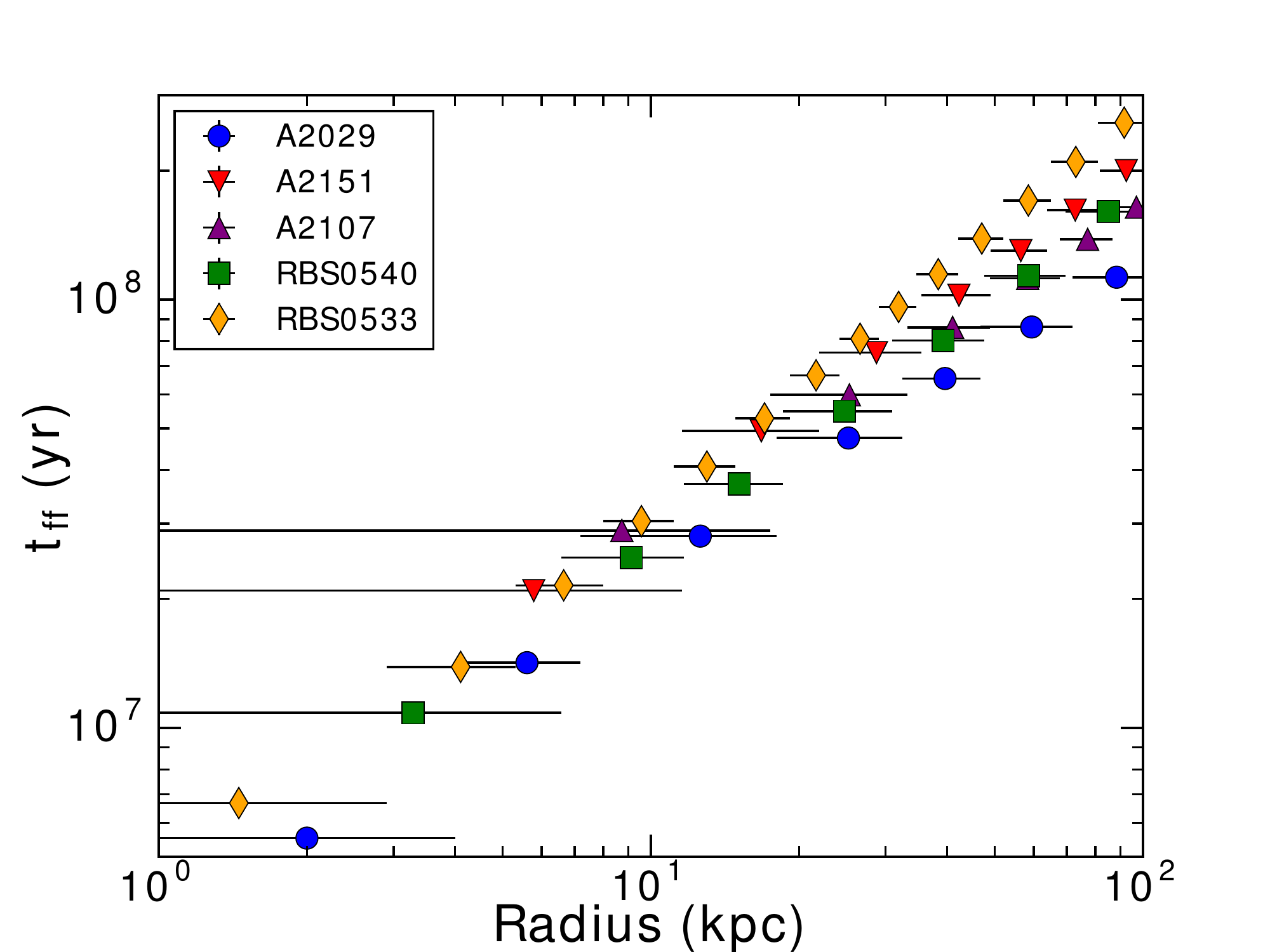}
		\includegraphics[width=0.46\linewidth, keepaspectratio]{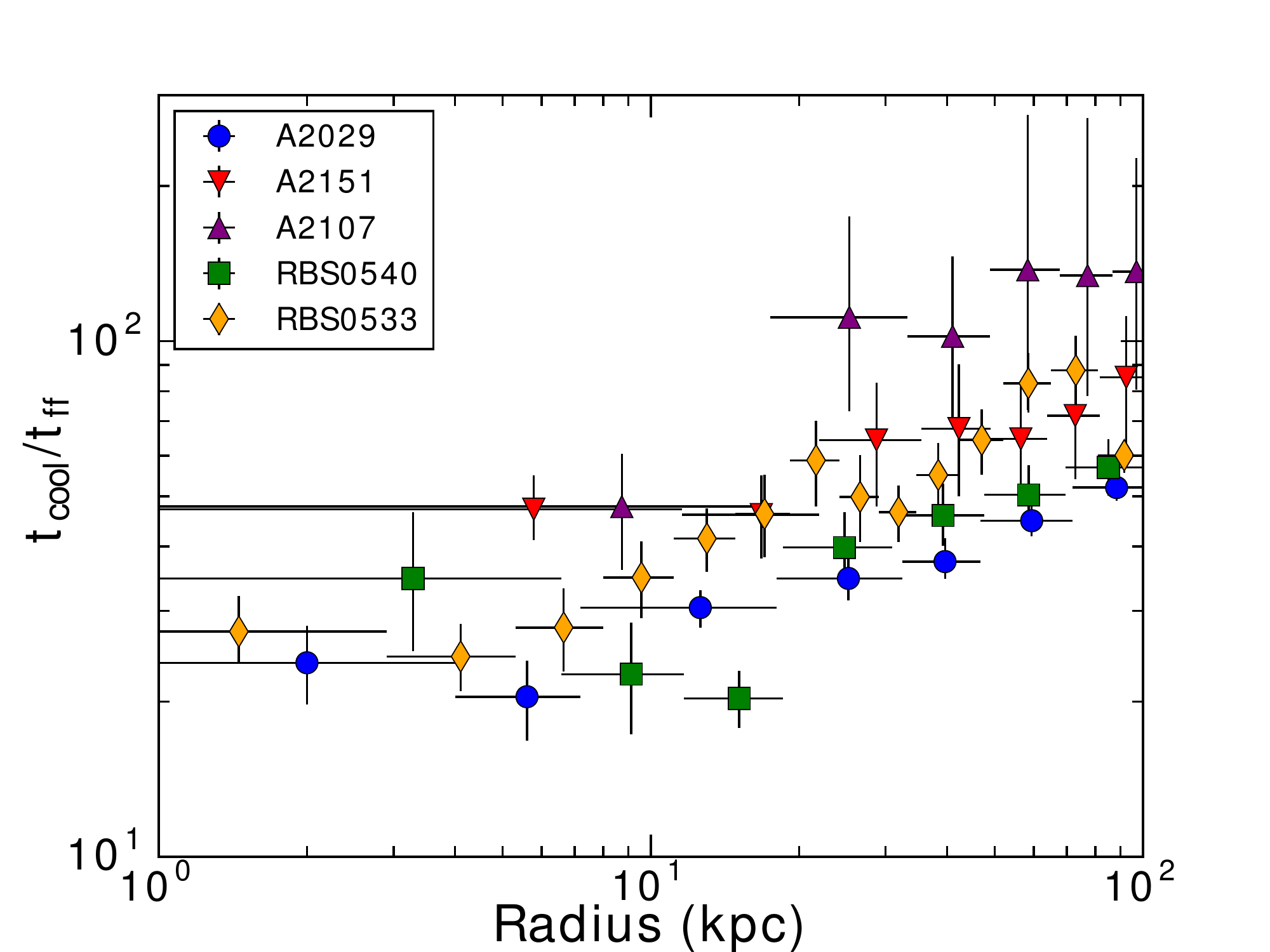}
		\caption{\textbf{Top-left}: The Enclosed mass found from mass fitting, see Table~\ref{Mass_Profile_Fits}. \textbf{Top-right}: Free-fall profiles calculated using Equation~\ref{tff}. \textbf{Bottom}: Deprojected $t_{\rm cool}$/$t_{\rm ff}$ profiles.}
		\label{Mass_tff_tcool/tff}
    \end{figure*}
    
	To compute the uncertainties in these quantities, \textsc{xspec} was used to create a Markov chain Monte Carlo (MCMC) simulation of 5000 iterations. We adopted the $1\sigma$ standard deviation as the uncertainties in our mass profiles as well as the uncertainties of $R_{s}$ and A$_{\rm NFW} = 4 \pi \mu G \rho R_{s}^{2}$. The total cluster mass can then be computed at $R_{2500}$, the radius where the mean density is 2500 times that of the critical density, $\rho_{c}$,
	\begin{equation}
	M_{2500} = \frac{4 \pi R_{2500}^{3}}{3}\overline{\rho},
	\end{equation}
	\noindent where $\bar{\rho}=2500 \rho_{c}$. The best fitting parameters are shown in Table~\ref{Mass_Profile_Fits}. The ratio of cooling time to free-fall time is believed to be related to thermally unstable cooling \citep{Nulsen_1986,Pizzolato&Soker_2005,McCourt_2012}, as such free-fall time profiles are derived for each cluster. The enclosed mass profiles obtained from fitting are used to calculate the local gravitational acceleration, $g=GM/R^{2}$, which can be used to calculate $t_{\rm ff}$: 
	\begin{equation}
	t_{\rm ff}(R) = \sqrt{\frac{2R}{g}}.
	\label{tff}
	\end{equation}
    \indent The enclosed cluster mass and free-fall time profiles are shown in the top-left and top-right of Figure~\ref{Mass_tff_tcool/tff}, respectively. The latter was used to create $t_{\rm cool}$/$t_{\rm ff}$ profiles shown in the bottom panel. The minimum $t_{\rm cool}$/$t_{\rm ff}$ values lie between $20-50$, with A2029 residing at the lower end of this range and A2107 being at the higher end. Our $\text{min}(t_{\rm cool}/t_{\rm ff})$ profiles differ from previous results, but agree within a $1\sigma$ error for values calculated for A2029 \citep{Hogan_2017a,McNamara_2016}, A2151 \citep{Pulido_2018}, and A2107 \citep{Hogan_2017a}. These differences are likely due to contrasts in the size and number of spatial bins used for spectral extraction.
    
	\section{Quantitative Analysis}
	\label{Quantative Analysis}
	Visual inspection of images such as Figure~\ref{Figure:BG_Sub_images} and Figure~\ref{Residual_Images} reveals two instances of sloshing swirls, in A2029 and A2151. The atmospheres of A2107 and RBS0540 are nearly structureless. Only RBS0533 has indications of surface brightness depressions consistent with an X-ray cavity or bubble. In this section, we present a method for estimating the significance of surface brightness depressions which can be used to determine whether or not these regions are X-ray cavities.
	
	\subsection{Surface Brightness Variations in the ICM}
	\label{SB Variations}
    \begin{figure*}
        \centering
        \includegraphics[width=90mm, height=90mm]{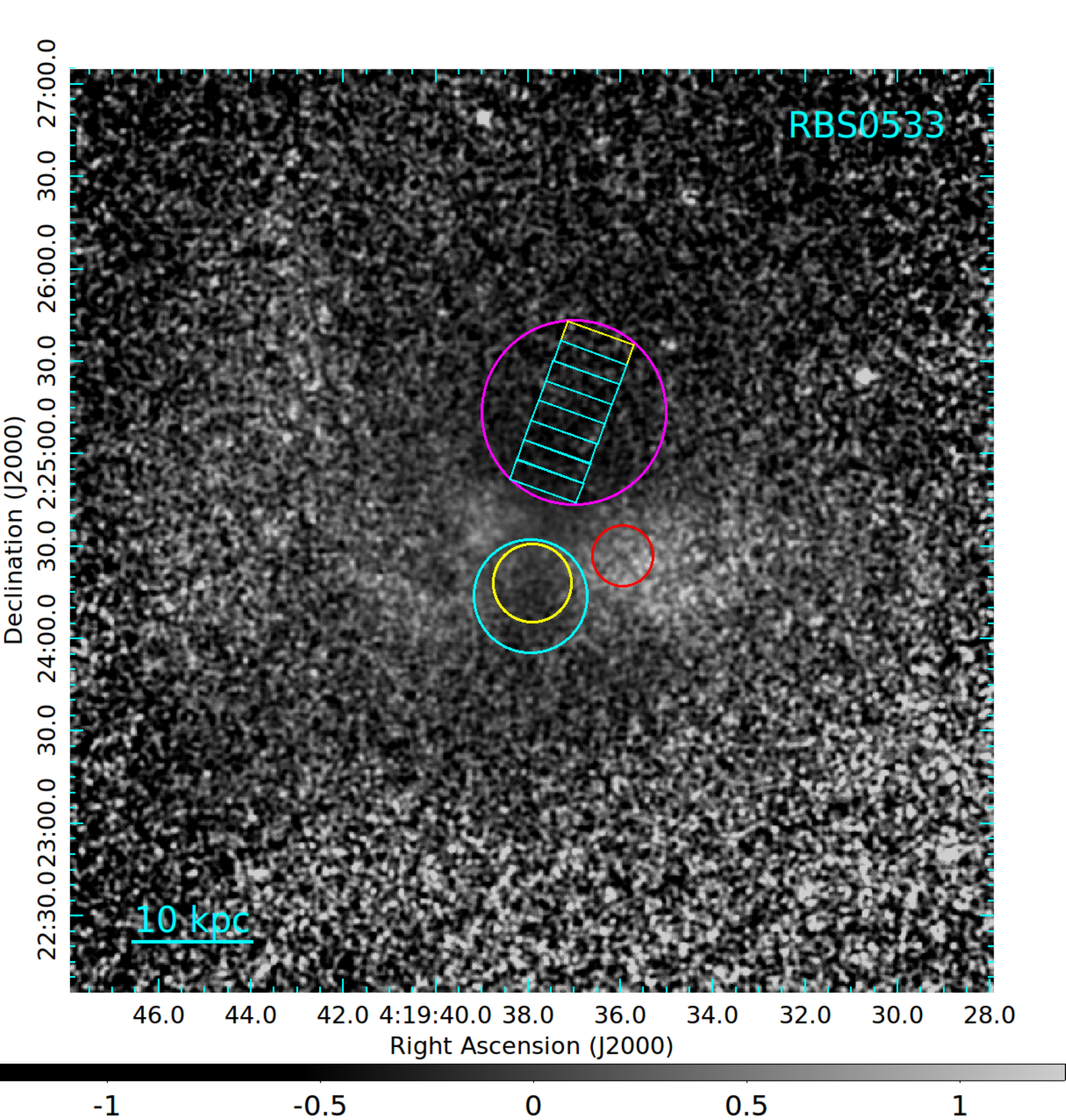}
        \caption{Model subtracted relative residual image of RBS0533 (same as in Figure~\ref{Residual_Images}) given by $(N_{I}-N_{M})$/$N_{M}$. Overlaid are circular regions used in significance testing, see Table~\ref{Table: Significance Test - Beta Model}, where the magenta circle is the approximate location and size of an X-ray bubble. Since the bubble has no clear rims, rectangular regions are used to find the approximate edge of a bubble where its structure is barely detected ($\rm SNR < 5$) which corresponds to the yellow rectangle, see text for more detail. Colour regions to the South of the magenta circle may also be a cavity, although the SNR within these regions is low. This image is Gaussian-smoothed with a 3 arcsec kernel radius.}
        \label{fig:RBS0533_Residuals}
    \end{figure*}	
	
	X-ray cavities are typically identified by surface brightness deficits of about $20-30\%$ relative to the surrounding medium \citep[]{McNamara_2007, McNamara_2012}. \citet{Panagoulia_2014}  showed that $\sim 20000$ counts within 20 kpc of the core is often required to clearly detect a cavity. Only A2029, RBS0533, and RBS0540 satisfy this criterion. \cite{Calzadilla_2018} studied the surface brightness fluctuations in A1664 to determine if two depressions surrounding the BCG are cavities or due to noise. Through significance tests, they determined that the regions were significant fluctuations, and thus cavities.
	
	We approached our analysis of surface brightness fluctuations by comparing the counts in the images of Figure~\ref{Figure:BG_Sub_images}, $N_{I}$, to the counts in the best fitting $\beta-$model image, $N_{M}$. The model represents the undisturbed cluster atmosphere. We used the residual images in Figure~\ref{Residual_Images} as a reference point for the location of potential bubbles in the ICM. We put circular regions of radius $r$ over these depressions at distance $R$ from the centre, and calculate the signal-to-noise ratio, SNR, in these regions to determine their significance. The signal within each generated region is calculated as,
	\begin{equation}
	|S| = |N_{I}-N_{M}|,
	\label{Signal}
	\end{equation}
	\noindent
	and the SNR within a region is estimated by,
	\begin{equation}
	\rm{SNR} = \frac{|S|}{\sqrt{|S| + 2 N_{\rm M}}}.
	\label{SNR}
	\end{equation}
	\indent We calculate N$_{M}$ by fitting a double $\beta-$model and single $\beta-$model to the surface brightness profiles with both elliptical and circular annuli centered on the brightest pixel. The residual images reveal one cluster, RBS0533, possesses bubble-like structure to the North of its centre. Since this bubble has no rims, making size estimates difficult, the size of the bubble was determined by calculating the SNR where the structure fades into the background (SNR $<5$). This was done by overlaying box regions with fixed length and width, corresponding to 5.7 kpc and 1.7 kpc, respectively. Regions are placed in succession of one another moving radially outwards, beginning at roughly 2.5 kpc where the depression is visible (see Figure~\ref{fig:RBS0533_Residuals}). We find that at approximately $R=20$ kpc (the edge of yellow the rectangle), the SNR falls below 5. Since the shape of the bubble is also unknown, we estimate the bubble as being spherically symmetric which encloses the boxed regions (magenta circle). 
	
	The SNR for the deficit in this circular region and for the other circular regions of interest are marked in Figure~\ref{fig:RBS0533_Residuals}. The best fit single and double $\beta-$model for elliptical and circular annuli are used to determine the deficits in these regions and the SNR is calculated within each region using Equation~\ref{SNR}. The details of this are given in Table~\ref{Table: Significance Test - Beta Model}.

	The surface brightness deficit of this candidate bubble reaches $31\%$ with a SNR of $\sim 28$. These values are consistent both with elliptical and circular $\beta$-models, indicating the structure is resilient to model parameters.  Thus it is likely real and is roughly consistent with the emissivity expected for an evacuated cavity relative to its surroundings \citep{McNamara_2007}.
	
	Coloured regions to the South of the centre (cyan and yellow circles) were found to be insignificant, and correspond to an excess of roughly 6\% at most for the single circular $\beta-$model and a deficit close to 2\% relative to the double elliptical $\beta-$model. The SNR is relatively low for all models in these regions, which suggests that the structure we see within these regions is likely not real, or rather, it may be an artifact from the model.
	
	The red circle to the West of the centre is a region of excess, which reaches levels of about $39\%$ and $47\%$ relative to the model corresponding to SNRs of $13$ and $15$ for an elliptical and circular double $\beta-$model, respectively. This indicates that this structure is likely a real feature.

	Extending this analysis to the other clusters, we find that no other clusters have significant structure associated with feedback, although, the sloshing feature in A2029 has a SNR $>20$ making it a conclusive detection, whereas the sloshing feature in A2151 has a SNR $\lesssim5$, making its detection uncertain.
	
	Finally, we can obtain an estimate for the total energy required to inflate the bubble in RBS0533, which is given by its enthalpy,
	\begin{equation}
	E = \frac{\gamma}{\gamma - 1}pV,
	\end{equation}
	\noindent where $p$ is the pressure within the cavity assuming that the cavity is in pressure balance with its surroundings, $V$ is the volume of the cavity, and $\gamma$ is the ratio of specific heat capacities. Here, $\gamma$ is 4/3 for a relativistic gas and 5/3 for a non-relativistic monatomic gas. Throughout our analysis, we assume that cavities are filled by a relativistic ideal gas, so $E=4pV$. The age of a cavity is best represented by the buoyancy time scale, $t_{\rm buoy}$, which is given by, \citep{Birzan_2004, Vantyghem_2014},
	\begin{equation}
	 t_{\rm buoy} \simeq R \sqrt{\frac{SC}{2gV}},
	\end{equation}
	\noindent where $S$ is the bubble's cross-section, $V$ is the volume of the bubble, $g$ is the local gravitational acceleration, and $C=0.75$ is the drag coefficient \citep{Churazov_2001}. The gravitational acceleration is estimated as $g=GM/R^{2}$, where $M$ is the total enclosed mass found in Section~\ref{Cluster Masses}. The buoyancy time of the bubble is $\sim 1.3\times 10^{7}$ yr which can be used to estimate the jet power, or mechanical power of the AGN that would be required to inflate a spherical bubble of this size, given by:
	\begin{equation}
	    P_{\rm jet} = \frac{4 pV}{t_{\rm buoy}}.
	\end{equation}
	\indent For a bubble of size $r=7.5$ kpc at distance $R=10$ kpc, the jet power is $3.5\pm0.2\times10^{43}$ erg s$^{-1}$.
	
    \begin{table*}[ht]
	\begin{center}
		\caption{Signal-to-noise ratio for regions in RBS0533 fit with an elliptical $\beta-$model and circular $\beta-$model.}
		\label{Table: Significance Test - Beta Model}
		\scalebox{1.0}{
		\begin{threeparttable}[whiteb]
			
			\begin{tabular}{@{}ccccccccccc}
				\hline\hline
				& & & \multicolumn{4}{c}{Elliptical model} & \multicolumn{4}{c}{Circular Model} \\ 
				\cmidrule(lr){4-7} \cmidrule(lr){8-11}
				Region Colour & R & r  & \multicolumn{2}{c}{Single $\beta-$model} & \multicolumn{2}{c}{Double $\beta-$model} & \multicolumn{2}{c}{Single $\beta-$model} & \multicolumn{2}{c}{Double $\beta-$model}\\ %
				\cmidrule(lr){4-5}\cmidrule(lr){6-7} \cmidrule(lr){8-9} \cmidrule(lr){10-11}
				& (kpc) & (kpc) & Deficit (\%) & SNR & Deficit (\%) & SNR &  Deficit (\%) & SNR & Deficit (\%) & SNR\\
				(1) & (2) & (3) & (4) & (5) & (4) & (5) & (4) & (5) & (4) & (5)   \\
				\Xhline{1.5\arrayrulewidth}
				\\
				Magenta & 10.1 & 7.5 & -29.4 & 26.0 & -31.0 & 27.6 & -28.5 & 25.1 & -30.0 & 26.6\\
				\Xhline{1.5\arrayrulewidth}
				\\
				Yellow & 4.5 & 3.2 & +2.5 & 1.6 & -2.4 & 1.6 & +5.1 & 3.1 & -0.2 & 0.1 \\
				\Xhline{1.5\arrayrulewidth}
				\\
				Cyan & 5.6 & 4.6 & +3.9 & 3.1 & -0.4 & 0.31 & +6.4 & 5.1 & +2.0 & 1.7  \\
				\Xhline{1.5\arrayrulewidth}
				\\
				Red & 8.6 & 2.5 & +47.3 & 14.8 & +39.2 & 12.8 & +54.6 & 16.4 & +47.2 & 14.8 \\
				\hline\hline
			\end{tabular}
			\begin{flushleft}\textbf{Columns}: (1) Region Colour in the Figure~\ref{fig:RBS0533_Residuals}, (2) Distance away from the cluster centre, (3) Size of region, (4) deficit of image relative to model ($N_{I}/N_{M} - 1$), (5) SNR given by Equation~\ref{SNR}.\end{flushleft}
		\end{threeparttable}}
	\end{center}		
    \end{table*}

	\begin{figure*}[!ht]
	 	
	\begin{center}
	 	\includegraphics[width=85mm, height=85mm]{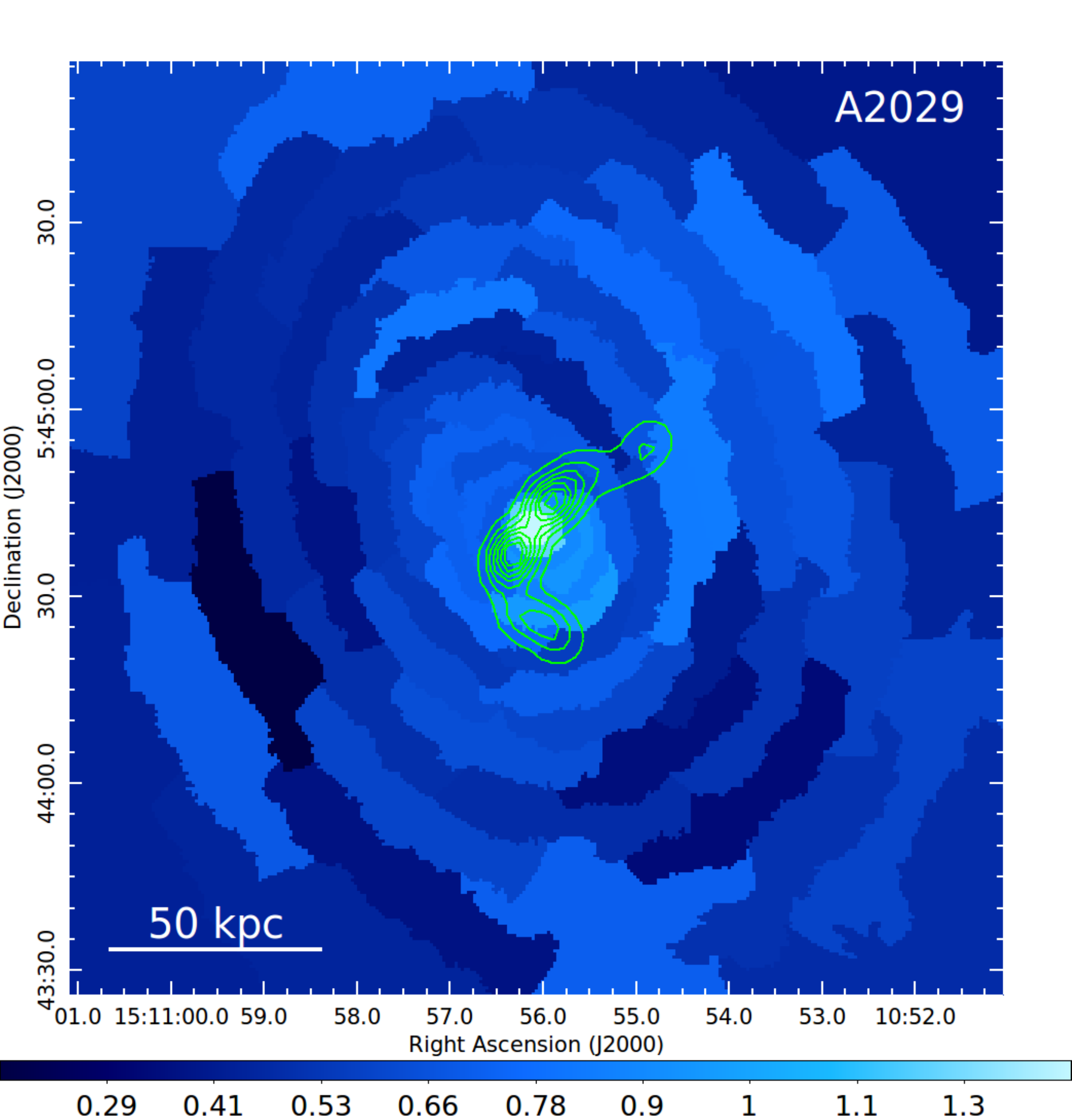}
	 	\includegraphics[width=85mm, height=85mm]{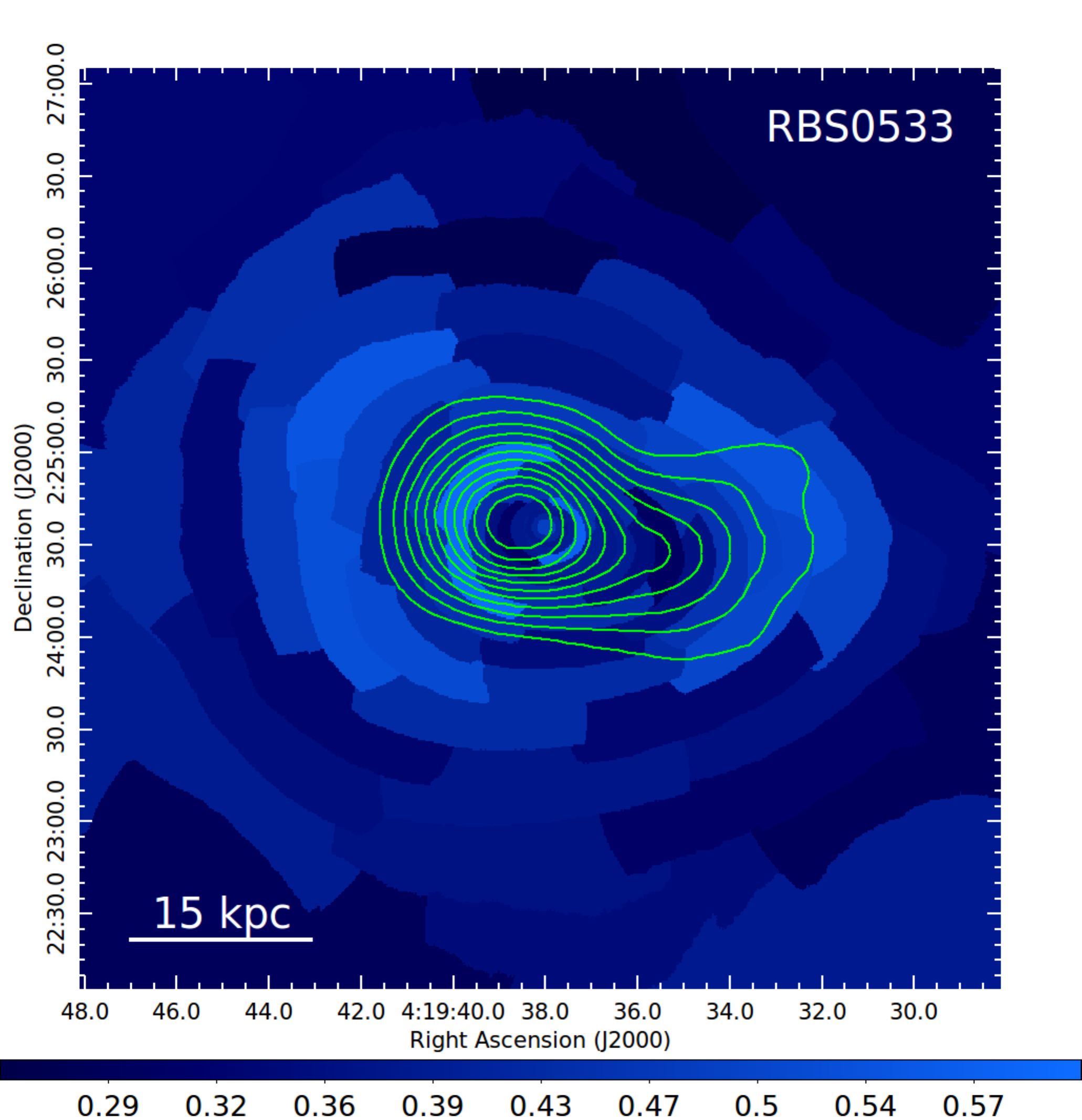}
	 	\includegraphics[width=85mm, height=85mm]{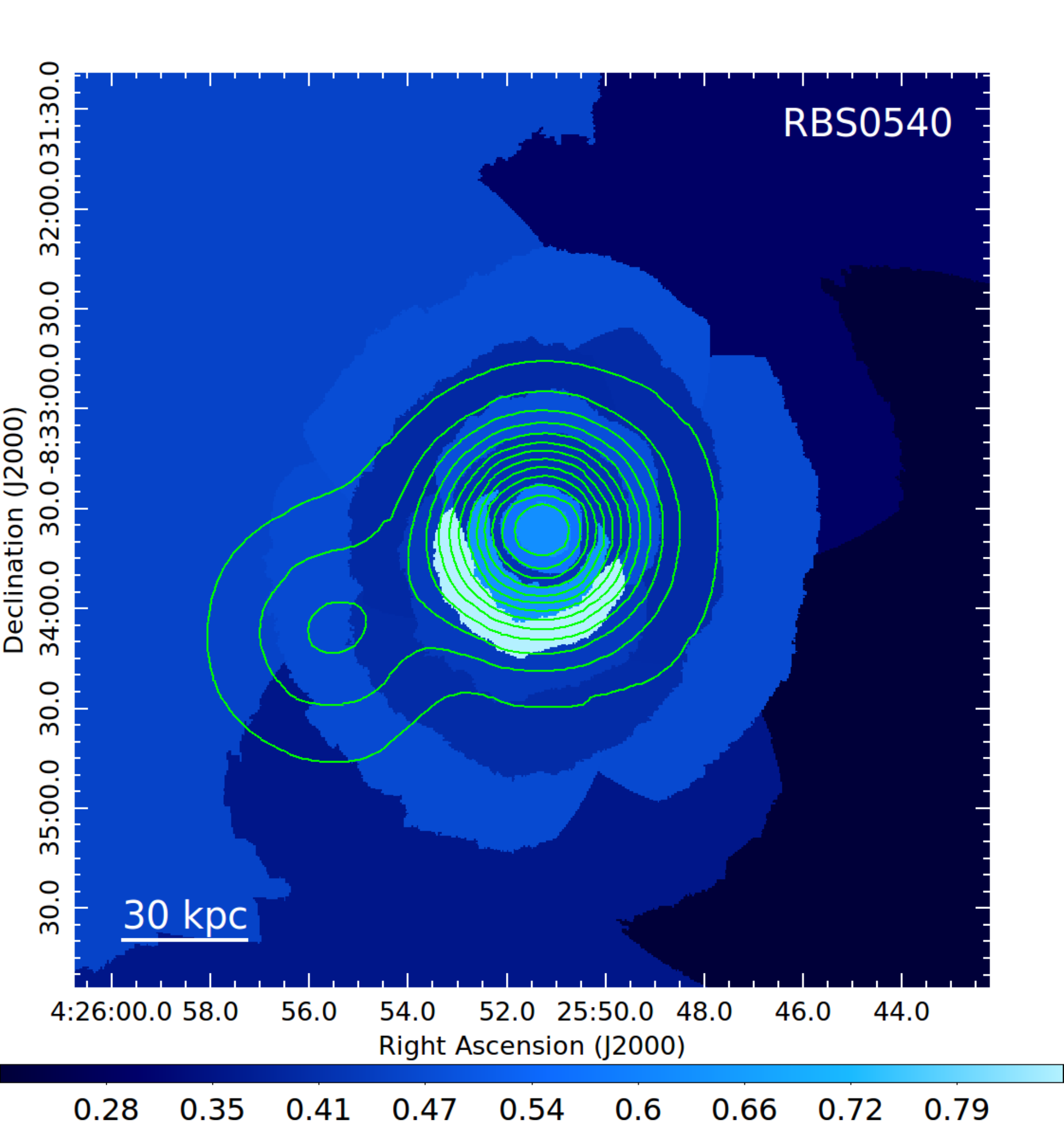}
	 	\includegraphics[width=85mm, height=85mm]{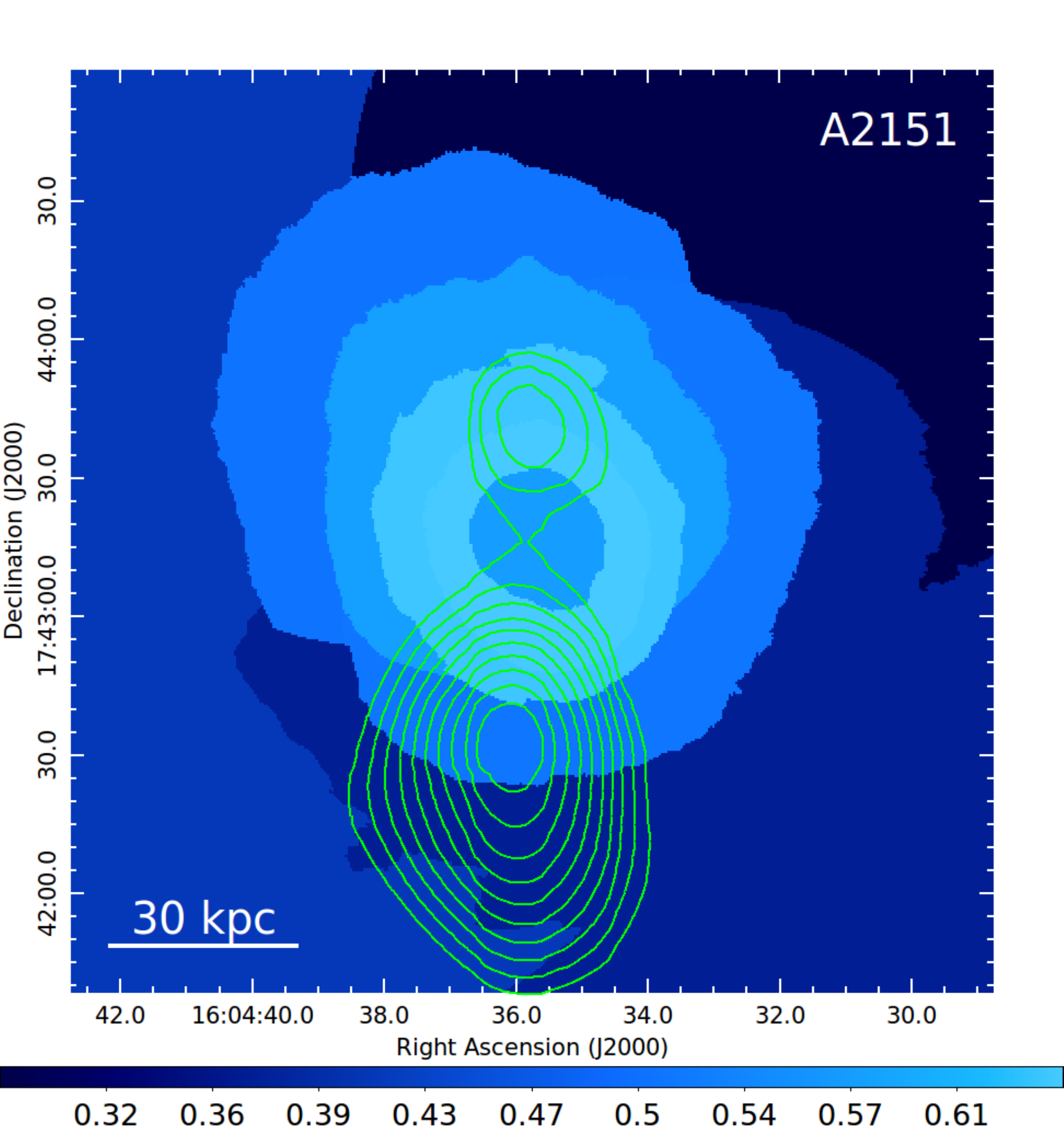}
	 	\caption{Metallicity maps of each cluster. Point sources were excluded from the images, where the colour bar is given in units of $Z_{\odot}$. The green contour lines are the radio data from the VLA FIRST survey shown at the 1.4 GHz frequency with a resolution of 5". In A2029, the radio observation has an rms noise of 11.8 mJy, with contours starting from 5$\sigma$ and increasing in steps of $0.001\times\sigma$. In RBS0533, the rms noise is 1.17 $\mu$Jy, with the contours beginning at 8$\sigma$ and increasing in steps of $0.5\times\sigma$. RBS0540's radio source has an rms noise at the 7.31 mJy level, with contours starting at 15$\sigma$ and increasing in steps of $0.1\times\sigma$. Lastly, A2151's radio source has an rms noise of the 1.63 mJy, with contours starting at 3$\sigma$ and increasing in steps of $0.2\times\sigma$. These maps show no evidence of metal-enriched plasma lying preferentially along jet axis.}
	 	\label{Maps}
    \end{center}
	 	
	\end{figure*}
    \subsection{Undetected Cavities}
    \label{Undetected Cavities}
    Among the factors that govern the detection of cavities, high central surface brightness favours their detection in cool cores. \cite{Birzan_2012} used a simulation to define the properties of bubbles that remain undetected. They concluded that most bubbles are undetected when the angle between the bubble-to-core axis and line of sight is small, or when they lie at large distances from the core.
    
    Apart from RBS0533, cavities may exist in these systems but remain undetected. Several studies have investigated the statistical properties of cluster cavities that are drawn from the \textit{Chandra} archive. One such survey found a cavity detection frequency of 41\% for a sample size of 75 clusters \citep{Birzan_2012}, while a sample of 133 systems biased towards cool core clusters found a detection rate of 52\% \citep{Shin_2016}. 
    
    In the brightest 55 clusters sample (B55), it was shown by \cite{Dunn&Fabian_2006} that 20 of these clusters require heating to offset cooling. At least 14 of the 20 clusters have clear bubbles and only one of these does not harbour a central radio source.
    
    Finally, \cite{Dunn_2008} studied the 42 clusters from the B55 and brightest cluster samples with \textit{Chandra} data. Of those, 23 have a central radio source. Defining cooling flow clusters as those with a significant central temperature drop and a short central cooling time, they found that 14 of the 42 clusters met these criteria and 6 of those harbour bubbles.
	
	The distribution of radio powers for this sample is consistent with expectations for cool core clusters \citep{Hogan_2015}. Although only one of the five objects in our sample has a significant surface brightness depression that may be an X-ray bubble, apart from A2107, four of five possess a central radio source shown in Figure~\ref{Maps}. It is noteworthy that the central radio source in RBS0533 does not coincide with the surface brightness depression associated with the putative X-ray bubble. While this does not exclude the possibility that it is a ``ghost cavity'' whose radio emission has faded \citep{Birzan_2004}, it is not a resounding confirmation that the surface brightness depression is indeed a radio/X-ray bubble. Our thermal instability analysis below assumes, conservatively, that it is indeed a bubble capable of lifting low entropy gas outward.
	
	\subsection{Uplifted Metal-Rich Atmospheric Gas}
	\label{Uplift}
	
	X-ray cavities not only displace hot gas, but may also draw metal-enriched plasma out from the centres of clusters at rates of tens to hundreds of solar masses per year \citep[]{Kirkpatrick_2011,McNamara_2014}. The maximum radius that metals can be uplifted to, referred to as the iron radius, $R_{\rm Fe}$, is correlated with $P_{jet}$ according to the relation \citep{Kirkpatrick_2015}:
	
	\begin{equation}
	\label{Iron_Radius}
	R_{\rm Fe} = (62 \pm 26) \times P_{\rm jet}^{0.45 \pm 0.06} \text{(kpc)}.
	\end{equation}

	Here $P_{\rm jet}$ is in units of $10^{44}$ erg s$^{-1}$ and $R_{\rm Fe}$ is defined as the radial bin furthest from the cluster's centre where the $1\sigma$ error bars for the metallicity profiles (along the jet and orthogonal to it) do not overlap \citep{Kirkpatrick_2011}. The scatter in the relation is large, approximately 0.87 dex. In clusters with known cavities, metal-enriched gas preferentially, but not exclusively, lies along the bubbles and thus the radio jets.
	
	\begin{figure}[h!]
		
		\begin{center}
			\includegraphics[height=51.6mm,keepaspectratio]{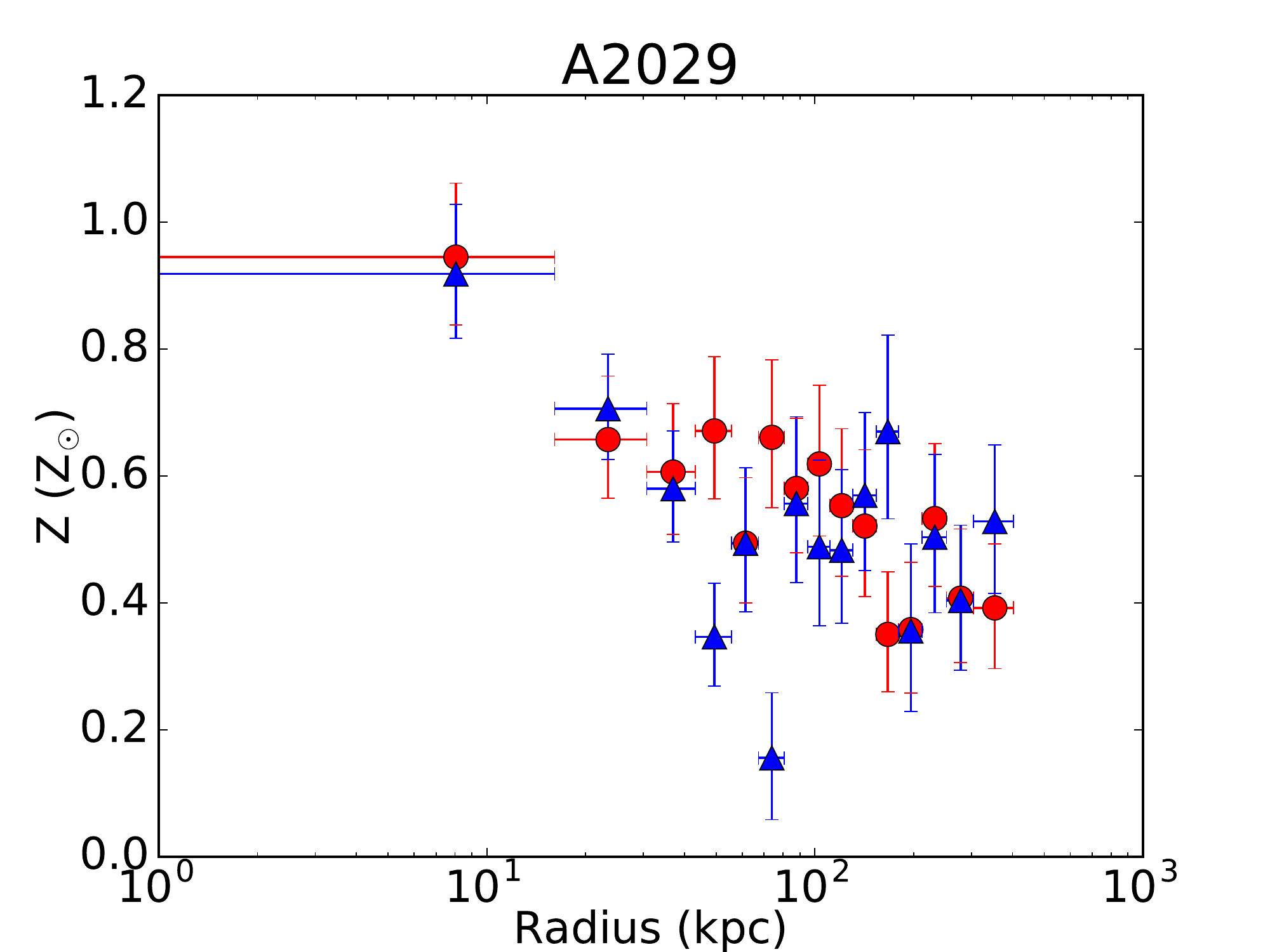}
			\includegraphics[height=51.6mm,keepaspectratio]{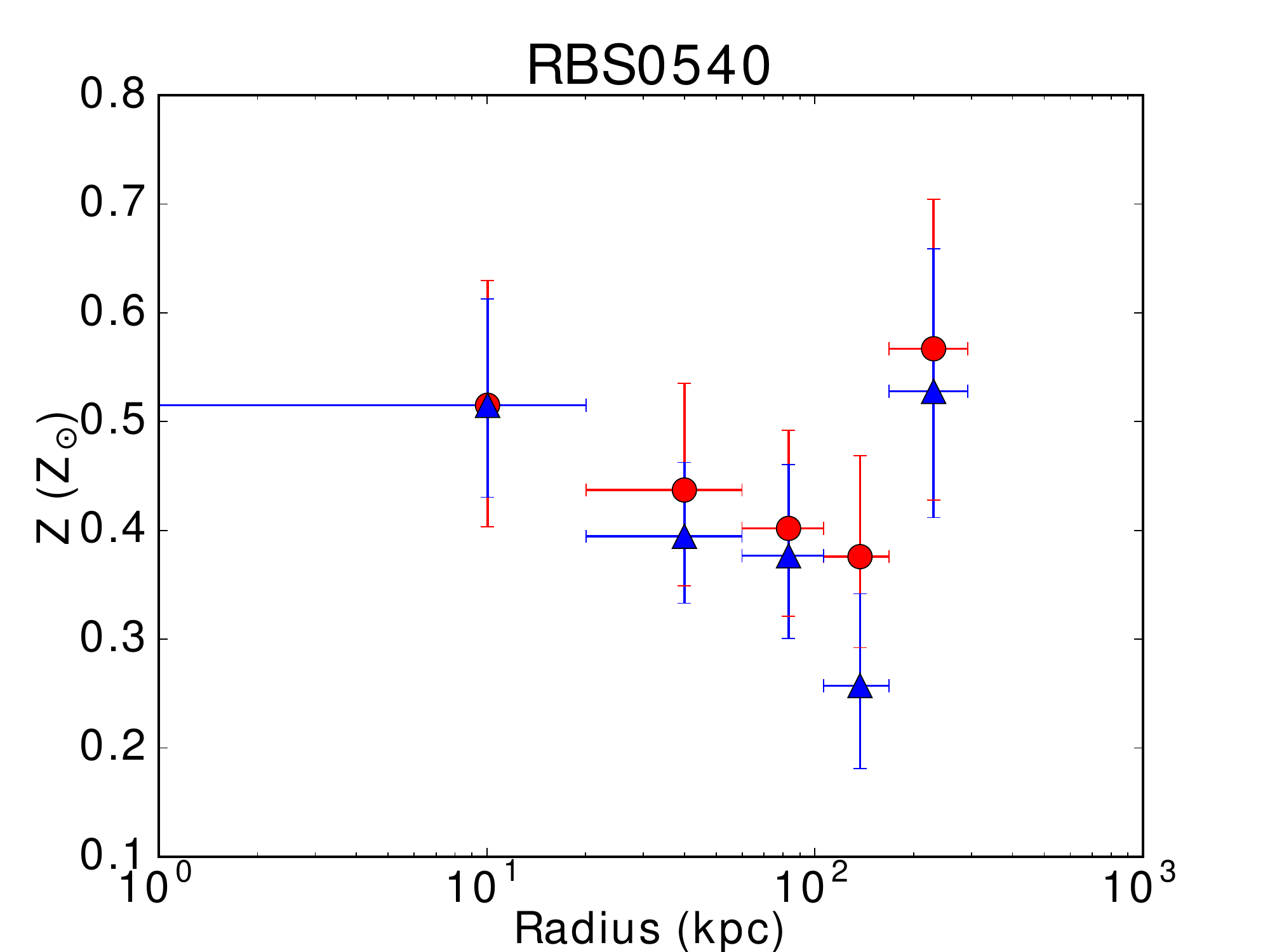}
			\includegraphics[height=51.6mm,keepaspectratio]{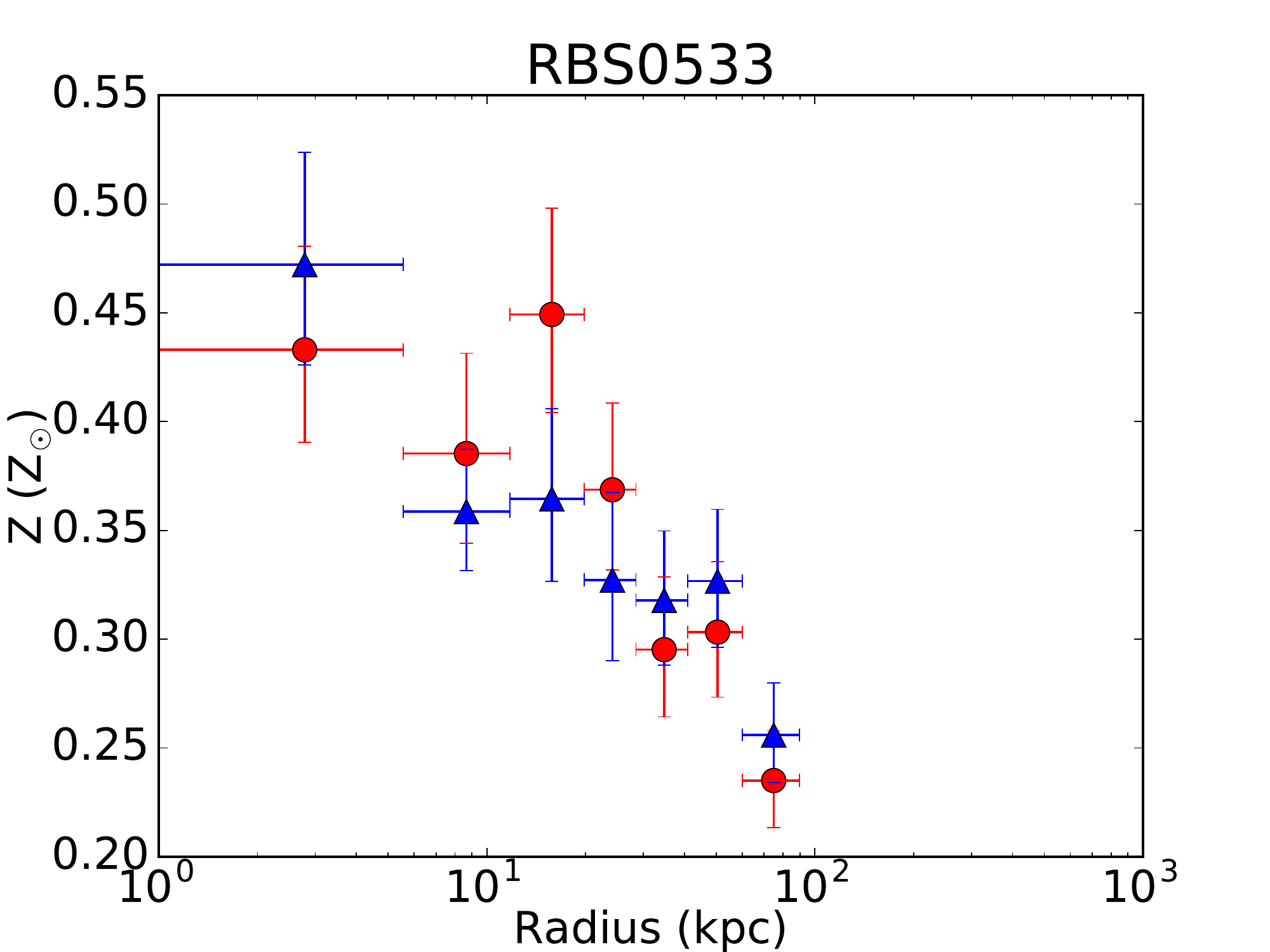}
			\includegraphics[height=51.6mm,keepaspectratio]{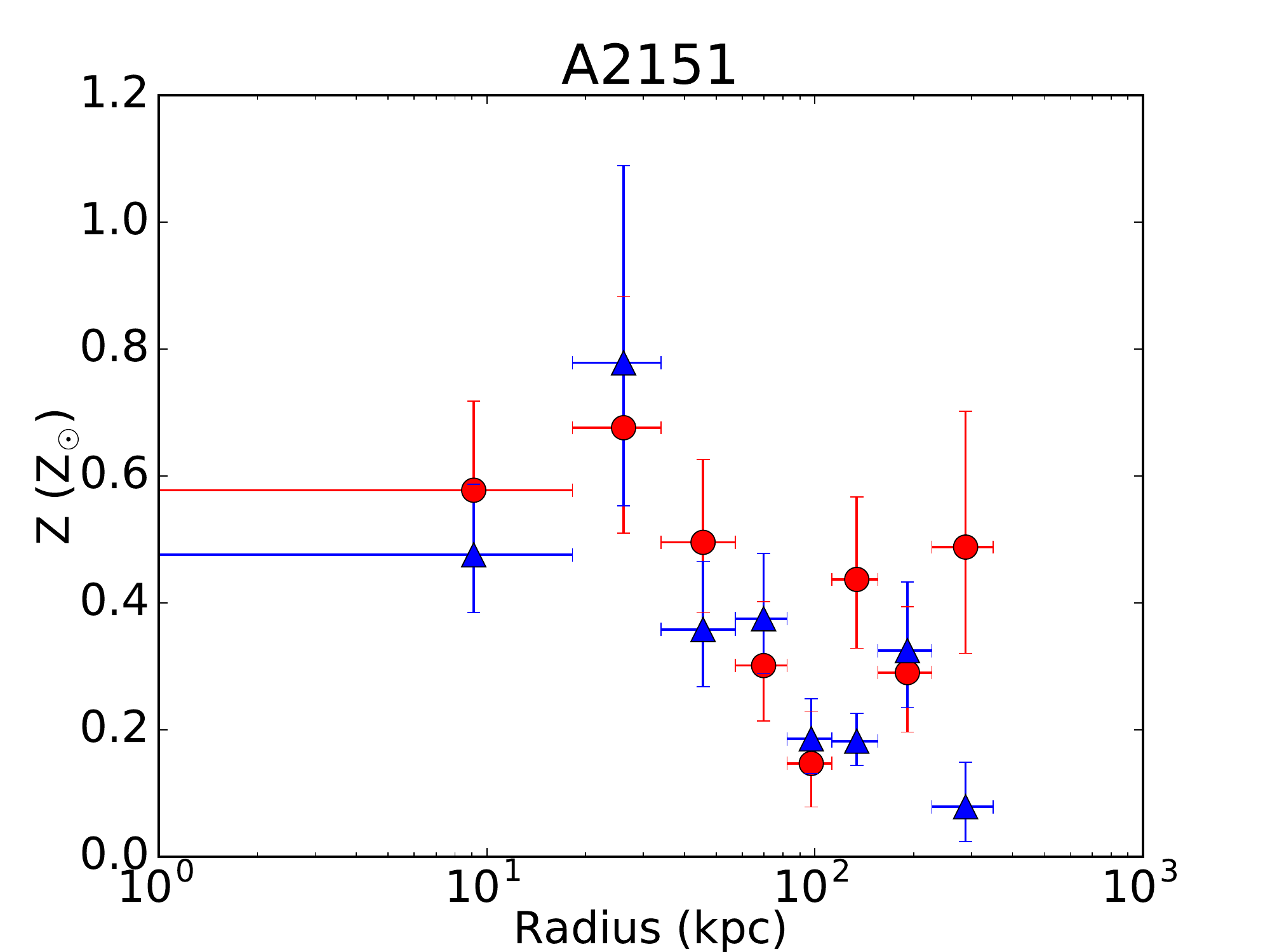}
			\caption{Metallicity profiles for the spoiler clusters, from top to bottom: A2029, RBS0540, RBS0533, A2151. Abundances along the jet axis are represented by red circles, while those for spectra in orthogonal directions or other off-jet locations are denoted by blue triangles. A2029's along-jet profile shown here is of its North-West jet. These profiles show no evidence that metal-enriched gas preferentially lies along the jet axis, which is consistent with Figure~\ref{Maps}. All errors here are reported at the 2$\sigma$ level.}
			\label{metallicity_jet_profiles}
		\end{center}
		
	\end{figure}
	
	\begin{table}[ht]
		\begin{center}
			\caption{Summary of \textit{t}-tests and KS-tests}
			\label{Table:Metallicity_Profiles_Stats}
			\begin{tabular}{@{}lcccc}
				\hline \hline
				\\
				& \multicolumn{2}{c}{\textit{t}-test} &  \multicolumn{2}{c}{KS-test} \\		
				\cmidrule(lr){2-3} \cmidrule(lr){4-5}
				Cluster  &  \textit{p}-value  &  Significant? & $D_{KS}$ & Same Sample?\\
				\hline
				\\
				A2029 & 0.52 & No & 0.0506 & Same \\ 
				\Xhline{1.5\arrayrulewidth}
				\\
				RBS0533 & 0.65 & No & 0.0191 & Same \\ 
				\Xhline{1.5\arrayrulewidth}
				\\
				RBS0540 & 0.40 & No & 0.146 & Same \\
				\Xhline{1.5\arrayrulewidth}
				\\
				A2151 & 0.42 & No & 0.168 & Same \\
				\hline \hline
			\end{tabular}
			Results of our \textit{t}-test and KS-test indicate no significant differences between along-jet and off-jet profiles in each of the clusters.
		\end{center}
	\end{table}		
	As a first step in a search for uplifted gas we created metallicity maps using the contour binning algorithm of \citet{Sanders_2006}. The metallicity maps are intended to provide a visual representation of the radial abundance distributions with respect to the radio sources. However, they are not used in our analytical evaluation of the relationship between the off-nuclear atmospheric gas abundance and the radio sources. Clusters were binned using a minimum signal-to-noise (S/N) of 70 per bin to maximize the number of bins generated while retaining high enough count per bin that uncertainties in metallicity do not dominate. No attempts were made to create maps for A2107 as its exposure time is too short, and thus its low number of counts would not allow us to generate enough bins with the required S/N for any meaningful analysis. 

	Spectra were extracted within each bin and fitted with a \textsc{phabs(apec)} model. Temperature, metallicity, and normalization were allowed to vary. Column densities were frozen at the value obtained from the LAB Survey \citep{Kalberla_2005}. The resulting metallicity maps are shown in Figure~\ref{Maps}.
	
    Higher metallicity gas aligned parallel and antiparallel to the jet axis is a strong indicator of metal-enriched gas being uplifted. The slightly asymmetric map near the centre of RBS0533 hints to this mechanism without being prominent enough to be deemed significant. 

	To further constrain the lack of evidence for uplifted metal-enriched gas in these four clusters, spectra were extracted from the profiles of annular sector bins with openings lying between 50 and 90 degrees. One of these profiles is along where the extended radio emission from the jet is located (``along-jet''). Extracted spectra in these regions are assumed to reflect the impact of the AGN on the gas. Other profiles are extracted along a direction orthogonal to or offset from the jet axis (``off-jet'') and represents the undisturbed atmosphere. Spectra from these profiles are assumed to be representative of the average prior AGN outbursts. Extracted spectra were fit in the same manner as those used to make Figure~\ref{Maps}. The results of this analysis are shown in Figure~\ref{metallicity_jet_profiles}. A2029 demonstrates slight evidence of a trend in higher metallicity gas along the jet-axis, but only in bins between approximately $20-60$ kpc outward from its centre.

	To determine if the profile along-jet and off-jet differ significantly, and thus the likelihood that an abundance excess along the jet axis is significant, we perform two statistical tests for each of our clusters. The results of both statistical tests are shown in Table~\ref{Table:Metallicity_Profiles_Stats}. 
	
	In our first test we compare the means for two profiles, using them to calculate the likelihood that they are drawn from the same distribution using a two-tailed \textit{t}-test. We choose to perform our tests at the 95\% significance level ($\alpha=0.05$) with the null hypothesis that differences between mean profile metallicities are insignificant. In each of the clusters, the null hypothesis cannot be rejected at the chosen significance level. 
	
	In the second statistical test, a two-sample Kolmogorov-Smirnov (KS) test is performed to determine if both the along-jet and off-jet metallicity profiles are independent of one another, or come from the same distribution. Again, a significance level of $\alpha=0.05$, is chosen with a null hypothesis, that both distributions are drawn from the same parent sample. The null hypothesis is rejected at this significance level if the calculated KS-statistic, $D_{KS}$, satisfies the condition given by $D_{KS} > \sqrt{-\frac{n+m}{2nm}\ln(\alpha)}$, where $n$ and $m$ are the sizes of the along-jet and off-jet profiles, respectively. Similarly to the previous test, the null hypothesis cannot be rejected in any of the clusters. 
	
	At the 95\% level then, there is no significant difference between metallicity profiles in any clusters, and thus there is no evidence of uplifted metal-enriched gas. Indeed, extracting spectra along random directions not aligned with the jet axis generally yield profiles that show no evidence of trends in our sample. Subsequent statistical analysis of our profiles only further confirms the results.

	\section{Discussion}
	\subsection{Thermally Unstable Cooling}
	\label{tcool-tff Discussion}
	  	
	 Central galaxies located at the bases of hot atmospheres are often associated with molecular clouds, star formation, and nebular emission. \textit{Chandra} observations have shown that they are prevalent when systems lie below the central cooling time and entropy thresholds (1 Gyr and 30 keV cm$^{2}$ respectively), while those above are usually devoid of cool gas and star formation \citep{Cavagnolo_2008, Rafferty_2008}.
     
	 On a more fundamental level, hot atmospheres should be susceptible to thermally unstable cooling when the ratio, $t_{\rm cool}/t_{\rm ff}$, falls below unity  \citep{Nulsen_1986,Pizzolato&Soker_2005,McCourt_2012}. In this context, the cooling time and entropy thresholds would be necessary but insufficient criteria.  However, the local value of $t_{\rm cool}/t_{\rm ff}$ almost never lies below 10, even in systems experiencing vigorous star formation \citep{Hogan_2017}. Others have suggested that thermally unstable cooling occurs when this ratio lies well above unity, in the range $10-30$ \citep{McCourt_2012, Sharma_2012, Gaspari_2012}.
	 
	 Inspection of Figures 4 \& 5 indicate that both the cooling time threshold and criterion are satisfied and that the $\text{min}(t_{\rm cool}/t_{\rm ff})\sim 20-50$. The central galaxies in this sample should be thermally unstable and should be forming stars and shining by nebular emission. They are not. This failure to respond to both criteria calls for a third possible criterion, possibly uplift, which we have investigated in detail here.  

	 \begin{table*}[ht!]
	    \begin{center}
		    \resizebox{0.99\textwidth}{!}{%
		    \begin{threeparttable}[ht]
		        \caption{Cold gas measurements for the spoiler clusters; A comparison of H$\alpha$ measurements from the ACCEPT database versus other sources.}
		        \label{Table: Cold Gas Measurements}
			    \begin{tabular}{@{}ccccccc}
				    \hline\hline
				    & & & & \multicolumn{2}{c}{Corrected}  &
				    \\ \cmidrule(lr){5-6}
				    Cluster & BCG & SFR$_{\rm UV}$ & L$_{\rm ACCEPT, H\alpha}$ & L$_{\rm H\alpha}$ & SFR$_{\rm H\alpha}$ & M$_{\rm H_{2}}$\\
			        &  & (M$_{\odot }$ yr$^{-1}$) & ($10^{40}$ erg s$^{-1}$) & ($10^{40}$ erg s$^{-1}$) & (M$_{\odot }$) & ($10^{8}$ M$_{\odot}$) \\
				    (1) & (2) & (3) & (4) &  (5) & (6) & (7) \\ 
				    \Xhline{1.5\arrayrulewidth}
				    \\
                    A2029 & IC 1101 &  $<1.72$ & $<0.643$ & $<0.44^{[1]}$ & $<0.035$ & $<17^{[2]}$ \\
                    \Xhline{1.5\arrayrulewidth}
				    \\
                    A2151 & NGC 6041 & $<0.38$ & $<0.141$ & $0.032^{[3]}$ & $0.003$ & $<3.1^{[2]}$\\
                    \Xhline{1.5\arrayrulewidth}
				    \\
                    A2107 & UGC 09958 & $<0.57$ & $<0.179$ & $-$ & $-$ & $-$ \\
                    \Xhline{1.5\arrayrulewidth}
				    \\
                    RBS0533 & NGC 1550 & $<0.14$ & $<0.016$ & $-$ & $-$ & $<0.47^{[4]}$ \\
                    \Xhline{1.5\arrayrulewidth}
				    \\
                    RBS0540 & MCG-01-12-005 & $0.4\pm0.09$ & $<0.011$ & $<0.014^{[5]}$ & $<0.001$ & $-$\\
				    \hline\hline
			    \end{tabular}%
			    \begin{flushleft}\textbf{Columns}: (1) Cluster, (2) BCG, (3) Ultraviolet SFR from \cite{Hoffer_2012}, (4) ACCEPT database H$\alpha$ luminosity \citep{Cavagnolo_2008}, (5) H$\alpha$ luminosity corrected for our chosen cosmology (see end of Section~\ref{Intro}), (6) SFR from calculated H$\alpha$ luminosity using $\rm{SFR}_{\rm H\alpha} = 7.9\times10^{-42} L_{\rm H\alpha}$ \citep{Kennicutt_1998}, (7) Molecular gas measurement from CO observations. \textbf{References} to $\rm{L_{\rm H\alpha}}$ and $\rm{M_{\rm H_{2}}}$ measurements: [1] \cite{McDonald_2010}, [2] \cite{Salome_2003}, [3] \cite{McDonald_2011}, [4] \cite{O'Sullivan_2018}, [5] \cite{Cavagnolo_2009}.  \end{flushleft}
		    \end{threeparttable}}
	    \end{center}		
     \end{table*}
     
	 We find surface brightness depressions consistent with an X-ray bubble only in RBS0533. However, it lacks bright rims composed of low entropy gas lifted from the inner region of the atmosphere. Furthermore, RBS0533 and the remaining clusters show no other evidence of substantial uplifted atmospheric gas that would trigger thermally unstable cooling once the gas reaches an altitude where $t_{\rm cool}/t_{\rm ff}$ falls below unity. Therefore, the observations are consistent with the hypothesis that uplift may be a significant factor in driving thermally unstable cooling. This investigation does not constitute proof, but may indicate we are on the right track. 
	 
	 Another factor that may trigger thermally unstable cooling is mild atmospheric turbulence \citep[]{Gaspari_2018, Voit_2018}. Turbulence may be induced by the peculiar motion of the central galaxy and mergers. However, in this context the driving mechanism would most likely be the central AGN. The absence of X-ray bubbles would imply the absence of a mechanism to drive the turbulence imparted on the lifted gas. The sloshing spiral in A2029 seen in Figure~\ref{Residual_Images} is evidence of a merger, and may indicate that it produced insufficient levels of turbulence required to trigger instabilities. 
     
	 On the other hand, modest atmospheric turbulence may be a factor leading to the thermal stability of these systems through turbulent heating \citep{Zhuravleva_2018, Hitomi_Perseus_2016}. The situation is far from clear but will be further explored with the X-ray Imaging and Spectroscopy Mission (\textit{XRISM}) and future X-ray observatories equipped with micro-calorimeter spectrometers.

	 \subsection{The Absence of Significant Cold Gas Mass}
	 \label{Cold Gas Discussion}
	 These objects were originally selected for observation on the basis of having upper limits on H$\alpha$ luminosity as listed in the ACCEPT database \citep{Cavagnolo_2008}. We have since performed an exhaustive literature search for more recent nebular and molecular mass measurements for each of the clusters.
	 
	 \subsection{RBS0533}
	 ACCEPT lists an H$\alpha$ luminosity for RBS0533 as $L_{\rm H\alpha}<0.016\times10^{40}$ erg s$^{-1}$. Two other studies that probed its central galaxy's (NGC 1550) CO emission, are in tension. \citet{O'Sullivan_2018} detected no CO(2-1) or CO(1-0) emission, arriving at an upper limit for molecular hydrogen of $\rm{M_{H2}} < 0.47\times 10^{8}$ M$_{\odot}$. However, \citep{Nakanishi_2007} claimed a detection of CO(3-2) deriving a molecular gas mass of $\rm{M_{H2}} = 4.3\times 10^{8}$ M$_{\odot}$. Clearly these measurements are inconsistent with each other. The apparent CO (3-2) line is broad, spanning a significant fraction of the receiver's $\sim 445$ km s$^{-1}$ bandwidth, leaving little room for the baseline continuum to be evaluated. Taking all into account, we adopt the \citep{O'Sullivan_2018} upper limit. RBS0533's central cooling time and entropy both lie below their respective thresholds ($10^{9}$ yr and $30$ keV cm$^{2}$), respectively and thus it is expected to shine in nebular emission.
	 
	 We have shown in Section~\ref{SB Variations} that its atmosphere harbors a possible X-ray bubble. Therefore, the absence of spatially-extended nebular emission is intriguing but not necessarily inconsistent with our hypothesis that thermally unstable cooling is stimulated by uplift. \cite{McNamara_2016} suggested that bubbles must lift cool, atmospheric gas to an altitude where $t_{\rm cool}/t_{\rm ff}<1$. The cooling time of the atmospheric gas at the centre of RBS0533 is $\sim 10^{8}$ yr. Based on Figure~\ref{Mass_tff_tcool/tff}, this gas must be lifted to an altitude of nearly $40$ kpc to meet this cooling criterion. However, the observed bubble, at least in projection, extends to roughly half this distance. Therefore, it is plausible the bubble has not lifted enough atmospheric gas to stimulate thermally unstable cooling at an observable level. 
     
     \subsection{A2151 \& A2107}
     A2151's H$\alpha$ luminosity is listed as $L_{\rm H\alpha}<0.141\times 10^{40}$ erg s$^{-1}$ in ACCEPT. \citet{McDonald_2011} recently etected H$\alpha$ emission in it's BCG, NGC 6041, with a luminosity of $L_{\rm H \alpha} \sim 3 \times 10^{38}$ erg s$^{-1}$. Therefore, NGC 6041 has a detectable level, albeit, a modest level of molecular gas. However, emission at this level lies well below the luminosity where the cooling time threshold seen in \cite{Cavagnolo_2008} becomes prominent at, $L_{\rm H \alpha} \sim 10^{40}$ erg s$^{-1}$. An H$\alpha$ detection of this magnitude is expected as the accumulation of gas from stellar winds, supernovae, and external accretion that may be unrelated to uplift and thermally unstable cooling. In A2107, we find no measurements for cold gas and so the ACCEPT upper limit is adopted.
    
    \subsection{RBS0540}
     RBS0540's H$\alpha$ luminosity, $<0.014\times10^{40}$ erg s$^{-1}$ \citep{Cavagnolo_2009}, agrees reasonably well with the value quoted in ACCEPT, $<0.011\times10^{40}$ erg s$^{-1}$. This is also the only cluster in our sample with a detection for SFR$_{\rm UV}$ of $0.4\pm0.09$ M$_{\odot}$ yr$^{-1}$ \citep{Hoffer_2012}.

     \begin{center}
     \begin{figure}[t]
        \centering
        \includegraphics[width=0.51\textwidth, keepaspectratio]{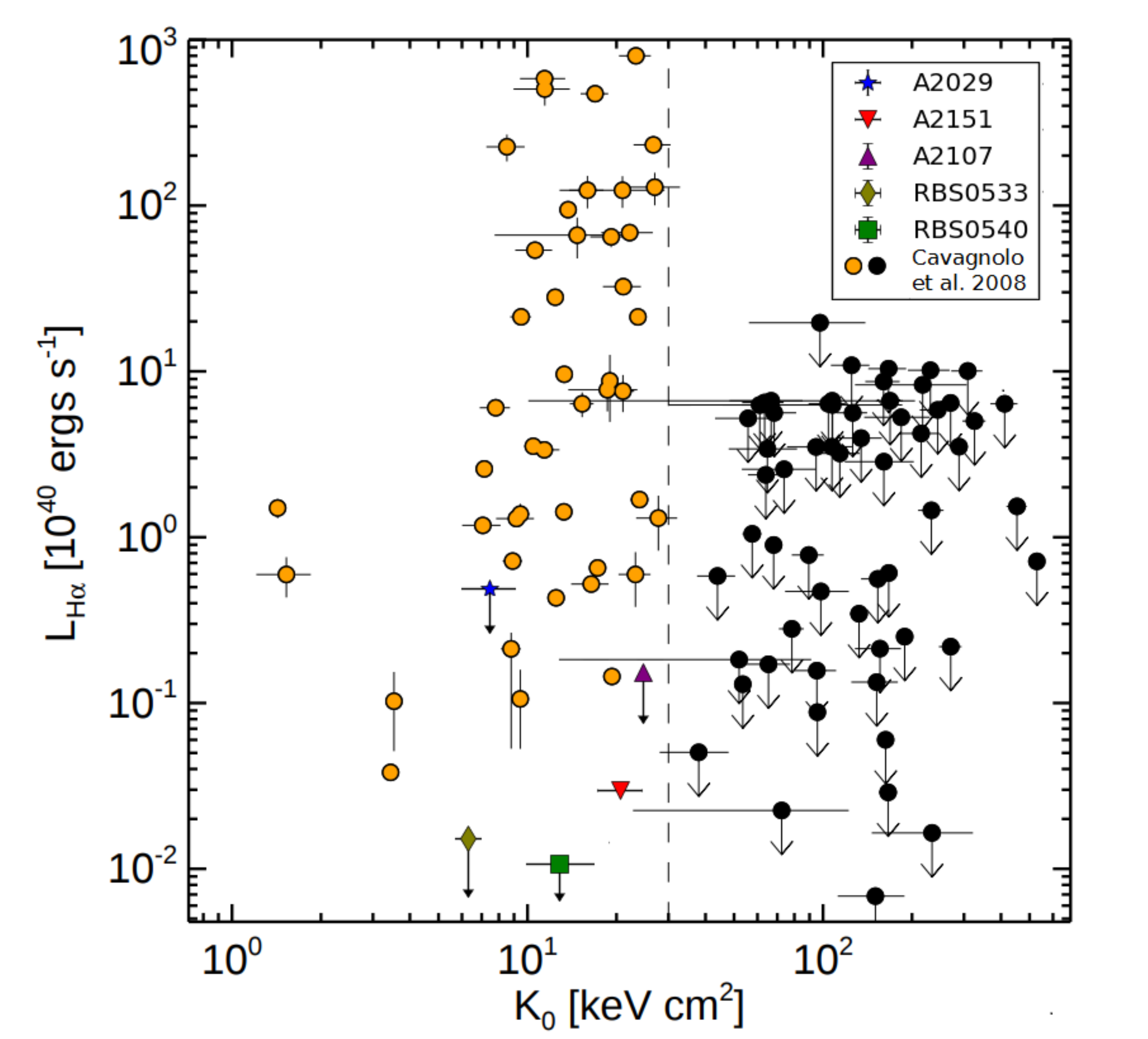}
        \caption{The central entropy ($R<10$ kpc) plotted against H$\alpha$ luminosity, adapted from \cite{Cavagnolo_2008}. Orange circles are H$\alpha$ detections, and black circles are non-detections of the clusters plotted against the H$\alpha$ luminosity. Overlaid are the points from our sample, all values are measured from column (5) in Table~\ref{Table: Cold Gas Measurements}, excluding RBS0533 and RBS0540, where the ACCEPT database values are plotted.}
        \label{fig:Halpha Luminosity vs min(entropy)}
     \end{figure}
     \end{center}

     \subsection{A2029}
     A2029's H$\alpha$ luminosity is listed in the ACCEPT database as $L_{\rm H\alpha}< 0.643 \times 10^{40}$ erg s$^{-1}$. However, \citet{McDonald_2010} found a more restrictive upper limit of $<4.41\times10^{39}$ erg s$^{-1}$. This upper limit is surprising as its central cooling time and bright, cuspy X-ray emission are similar to clusters with nebular emission luminosities exceeding this limit by more than three orders of magnitude.  Why A2029's central galaxy lies dormant while others with similar or less extreme atmospheric properties burgeon with star formation has been a mystery for decades.
     
     Further complicating the matter, A2029's central galaxy hosts a large and relatively powerful radio source $P_{1.4}\sim 10^{41}$ erg s$^{-1}$. Therefore, it should in principle be able to lift hot gas outward making it susceptible to thermally unstable cooling. 
     However, no evidence for large cavities was found in A2029 by \citet{Paterno-Mahler_2013} or in our analysis. A2029's radio source is long and thin (Figure~\ref{Maps}). It lacks jets feeding high-volume lobes seen in other powerful sources such as Hydra A and MS0735+096, which are lifting vast quantities of atmospheric gas \citep{Kirkpatrick_2015}.  It is unclear why this is so. It may be a consequence of atmospheric sloshing that may be sweeping the radio source back into a wide angle tail morphology \citep{Paterno-Mahler_2013}.
     
     In the context of this discussion, the absence of prominent X-ray cavities or radio lobes may indicate that its radio source is incapable of lifting an appreciable amount of atmospheric gas. Its mechanical power may be too small despite its powerful synchrotron emission. Therefore, its atmosphere remains thermally stable, at least for the time being. It is unclear why this would be. \citet{Croston_2018} have pointed out that Fanaroff \& Riley class (FR) I and II radio galaxies have different particle contents, with FRIs having higher jet (mechanical) power for their synchrotron luminosities than FRIIs. Perhaps A2029's radio source is composed of light particles akin to an FRII radio source rather than a FRI which is commonly found at the centres of clusters.
     
	 \section{Conclusions}
	 \label{Conclusion}
	 
	 In this paper we have studied five galaxy clusters using \textit{Chandra} observations and archival data. The clusters were selected from the ACCEPT database on the basis of possessing an upper limit on nebular $H\alpha$ emission. Our main findings are:
	 \begin{enumerate}
	 
	 	\item{Projected and deprojected thermodynamic profiles reveal that within the central 10 kpc of each cluster the atmospheric cooling time and entropy lie below $10^{9}$ yr, and $30$ keV cm$^{2}$, respectively. Below these thresholds, cool gas and star formation traced by nebular emission above $\simeq 10^{41} \rm ~erg~s^{-1}$ is commonly observed (Figure~\ref{fig:Halpha Luminosity vs min(entropy)}).}
	 	
	 	\item {Only RBS0533, has atmospheric structure consistent with a possible X-ray cavity. The feature is a 31\% depression in Figure~\ref{Figure:BG_Sub_images} relative to the elliptical double $\beta-$model, with a SNR of $\sim 28$.} 
	 	
	 	\item{While only one of five targets contain at least one cavity, four of the five clusters have radio emission as shown in Figure~\ref{Maps}. This property is consistent with other systems with short central cooling times \citep{Cavagnolo_2008}. The central galaxy in Abell 2151 possesses weak H$\alpha$ emission at the level of $3\times 10^{38}\rm ~erg ~s^{-1}$.  This level lies roughly 300 times below the level normally associated with cluster cooling. The absence of a significant levels of cold gas is consistent with the hypothesis these objects are able to effectively lift low entropy gas to an altitude where the atmosphere becomes thermally unstable, i.e., $t_{\rm cool}/t_{\rm ff}\lesssim 1$.}
	 	
	 	\item {Thermodynamic profiles extracted along and off the jet axis shows no evidence of uplift or that higher metallicity gas lies preferentially along the jet-axis. This is clearly evident in our abundance maps within Figure~\ref{Maps} and Figure~\ref{metallicity_jet_profiles}, and was confirmed more rigorously through statistical analyses using a \textit{t}-test and KS-test.}
	\end{enumerate}
	
\bigskip	
    We thank Jeremy Sanders and Keith Arnaud for their support in troubleshooting software problems with contour binning and Xspec, respectively. MTH, and BRM acknowledge funding from the Chandra X-ray Observatory Cycle 18 proposal for A2151 and RBS0540. HRR acknowledges support from an STFC Ernest Rutherford Fellowship and an Anne McLaren Fellowship.

	\bibliographystyle{apj}
	\bibliography{references}
	
	 

\end{document}